\documentclass[sigconf]{acmart}

\usepackage{amsmath,amsfonts}
\usepackage{graphicx}
\usepackage{textcomp}
\usepackage{bbding}
\usepackage{pifont}
\usepackage{multicol}

\usepackage{booktabs}
\usepackage[binary-units=true]{siunitx}
\usepackage{array,multirow}
\usepackage{hyperref}
\usepackage{amsmath}
\usepackage{enumitem}
\usepackage{graphicx}
\usepackage{float}
\usepackage[symbol, hang,flushmargin]{footmisc}
\usepackage{bm}
\usepackage{subfigure}

\usepackage{tablefootnote}

\usepackage[noend]{algpseudocode}

\usepackage[ruled,linesnumbered]{algorithm2e}
\SetKwComment{Comment}{$\triangleright$\ }{}
\newlength\mylenin

\let\oldnl\nl 
\newcommand{\nonl}{\renewcommand{\nl}{\let\nl\oldnl}} 

\newlength\mylenout

\newcommand{\figref}[1]{Figure~\ref{#1}}
\newcommand{\secref}[1]{Section~\ref{#1}}

\newcommand{\tabref}[1]{Table~\ref{#1}}

\newcommand{\Hsection}[1]{\vspace{0.5\baselineskip}\par\noindent\textit{#1}~\textbf{---}~}

\DeclareMathOperator*{\argmin}{argmin}

\newcolumntype{L}[1]{>{\raggedright\let\newline\\\arraybackslash\hspace{0pt}}m{#1}}
\newcolumntype{C}[1]{>{\centering\let\newline\\\arraybackslash\hspace{0pt}}m{#1}}
\newcolumntype{R}[1]{>{\raggedleft\let\newline\\\arraybackslash\hspace{0pt}}m{#1}}

\usepackage{soul}
\usepackage{xurl}
\usepackage{xspace}

\newcommand{\sys}{\textit{Dysta}\xspace}

\newif\ifarxiv
\arxivtrue
\usepackage{tikz}

\usepackage{fancyhdr}
\fancypagestyle{firstpage}
{
    \fancyhead{}
    \begin{tikzpicture}[remember picture,overlay]
    \node [xshift=134mm,yshift=-10mm]
    at (current page.north west) {\href{https://www.acm.org/publications/policies/artifact-review-and-badging-current}{\includegraphics[width=1.8cm]{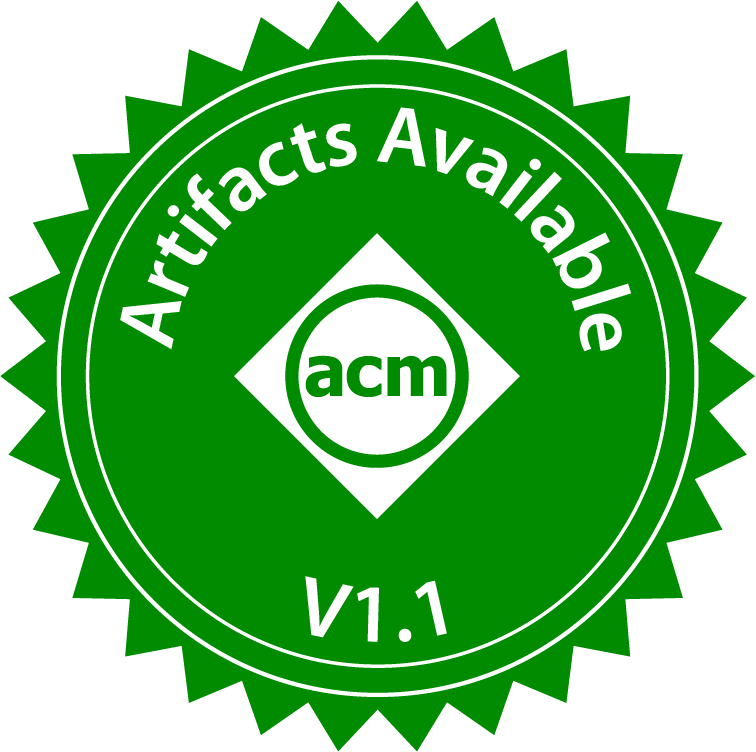}}} ;
    \node [xshift=153mm,yshift=-10mm]
    at (current page.north west) {\href{https://www.acm.org/publications/policies/artifact-review-and-badging-current}{\includegraphics[width=1.8cm]{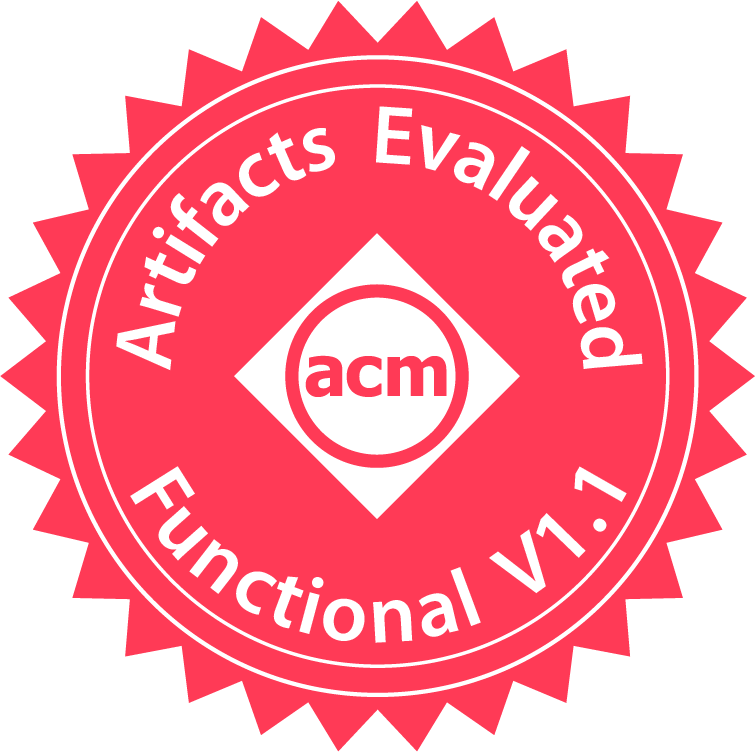}}} ;
    \node [xshift=172mm,yshift=-10mm]
    at (current page.north west) {\href{https://www.acm.org/publications/policies/artifact-review-and-badging-current}{\includegraphics[width=1.8cm]{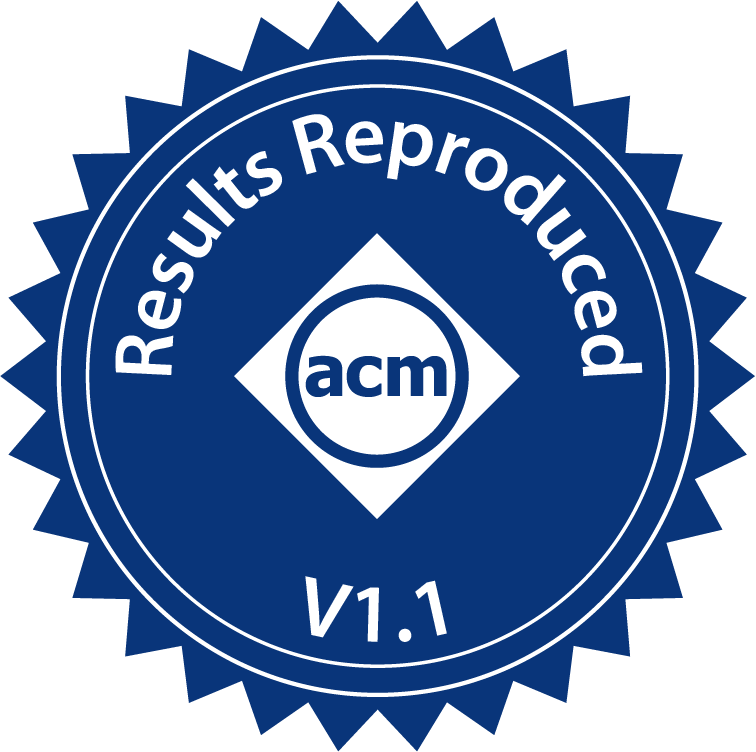}}} ;
    \node [xshift=191mm,yshift=-10mm]
    at (current page.north west) {\href{https://www.acm.org/publications/policies/artifact-review-and-badging-current}{\includegraphics[width=1.8cm]{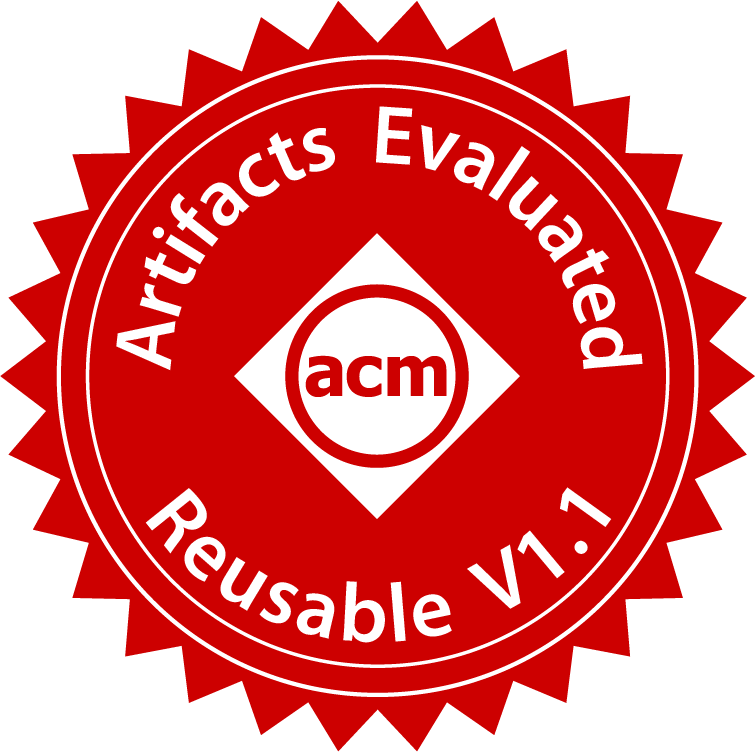}}} ;
    \end{tikzpicture}

}

\makeatletter
\def\ps@IEEEtitlepagestyle{%
  \def\@oddfoot{\mycopyrightnotice}%
  \def\@oddhead{\hbox{}\@IEEEheaderstyle\leftmark\hfil\thepage}\relax
  \def\@evenhead{\@IEEEheaderstyle\thepage\hfil\leftmark\hbox{}}\relax
  \def\@evenfoot{}%
}
\def\mycopyrightnotice{%
  \begin{minipage}{\textwidth}
  \scriptsize
  \copyright~2021 IEEE. Personal use of this material is permitted. Permission from IEEE must be obtained for all other uses, in any current or future media, including reprinting/republishing this material for advertising or promotional purposes, creating new collective works, for resale or redistribution to servers or lists, or reuse of any copyrighted component of this work in other works. 
  
  This work has been accepted at The International Conference on Field-Programmable Technology (FPT’21).
  \end{minipage}
}
\makeatother

\AtBeginDocument{%
  }

\copyrightyear{2023}
\acmYear{2023}
\setcopyright{acmlicensed}\acmConference[MICRO '23]{56th Annual IEEE/ACM International Symposium on Microarchitecture}{October 28-November 1, 2023}{Toronto, ON, Canada}
\acmBooktitle{56th Annual IEEE/ACM International Symposium on Microarchitecture (MICRO '23), October 28-November 1, 2023, Toronto, ON, Canada}
\acmPrice{15.00}
\acmDOI{10.1145/3613424.3614263}
\acmISBN{979-8-4007-0329-4/23/10}

\begin{document}

\title{Sparse-DySta: Sparsity-Aware Dynamic and Static Scheduling for Sparse Multi-DNN Workloads}

\author{Hongxiang Fan}
\email{hongxiangfan@ieee.org}
\orcid{0000-0003-2387-5611}
\affiliation{%
  \institution{Samsung AI Center \& \\ University of Cambridge}
  \city{Cambridge}
  \country{UK}
}

\author{Stylianos I. Venieris}
\authornote{Corresponding author.}
\email{s.venieris@samsung.com}
\orcid{1234-5678-9012}
\affiliation{%
  \institution{Samsung AI Center}
  \city{Cambridge}
  \country{UK}
}

\author{Alexandros Kouris}
\email{a.kouris@samsung.com}
\orcid{1234-5678-9012}
\affiliation{%
  \institution{Samsung AI Center}
  \city{Cambridge}
  \country{UK}
}

\author{Nicholas D. Lane}
\email{ndl32@cam.ac.uk}
\orcid{0000-0003-2387-5611}
\affiliation{%
  \institution{University of Cambridge \& \\Flower Labs }
  \city{Cambridge}
  \country{UK}
}

\renewcommand{\shortauthors}{Hongxiang Fan et al.}

\begin{abstract}
Running multiple deep neural networks (DNNs) in parallel has become an emerging workload in both edge devices, such as mobile phones where multiple tasks serve a single user for daily activities, and data centers, where various requests are raised from millions of users, as seen with large language models.
To reduce the costly computational and memory requirements of these workloads, various efficient sparsification approaches have been introduced, resulting in widespread sparsity across different types of DNN models. 
In this context, there is an emerging need for scheduling sparse multi-DNN workloads, a problem that is largely unexplored in previous literature. 
This paper systematically analyses the use-cases of multiple sparse DNNs and investigates the opportunities for optimizations.
Based on these findings,
we propose \textit{Dysta}, a novel bi-level dynamic and static scheduler that utilizes both static sparsity patterns and dynamic sparsity information for the sparse multi-DNN scheduling.
Both static and dynamic components of \textit{Dysta} are jointly designed at the software and hardware levels, respectively, to improve and refine the scheduling approach.
To facilitate future progress in the study of this class of workloads,
we construct a public benchmark that contains sparse multi-DNN workloads across different deployment scenarios, spanning from mobile phones and AR/VR wearables to data centers.
A comprehensive evaluation on the sparse multi-DNN benchmark demonstrates that our proposed approach outperforms the state-of-the-art methods with up to $10$\% decrease in latency constraint violation rate and nearly $4\times$ reduction in average normalized turnaround time.
Our artifacts and code are publicly available
at: \url{https://github.com/SamsungLabs/Sparse-Multi-DNN-Scheduling}.
\end{abstract}

\ccsdesc[500]{Computing methodologies~Neural networks}
\ccsdesc[500]{Theory of computation~Scheduling algorithms}
\ccsdesc[500]{Computer systems organization~Real-time system architecture}

\keywords{Sparse Multi-DNN Scheduling, Dynamic and Static Approach, Algorithm and Hardware Co-Design}


\maketitle
\thispagestyle{firstpage}

\section{Introduction}\label{sec:intro}

\begin{figure}[t]\centering
\includegraphics[width=0.49\textwidth]{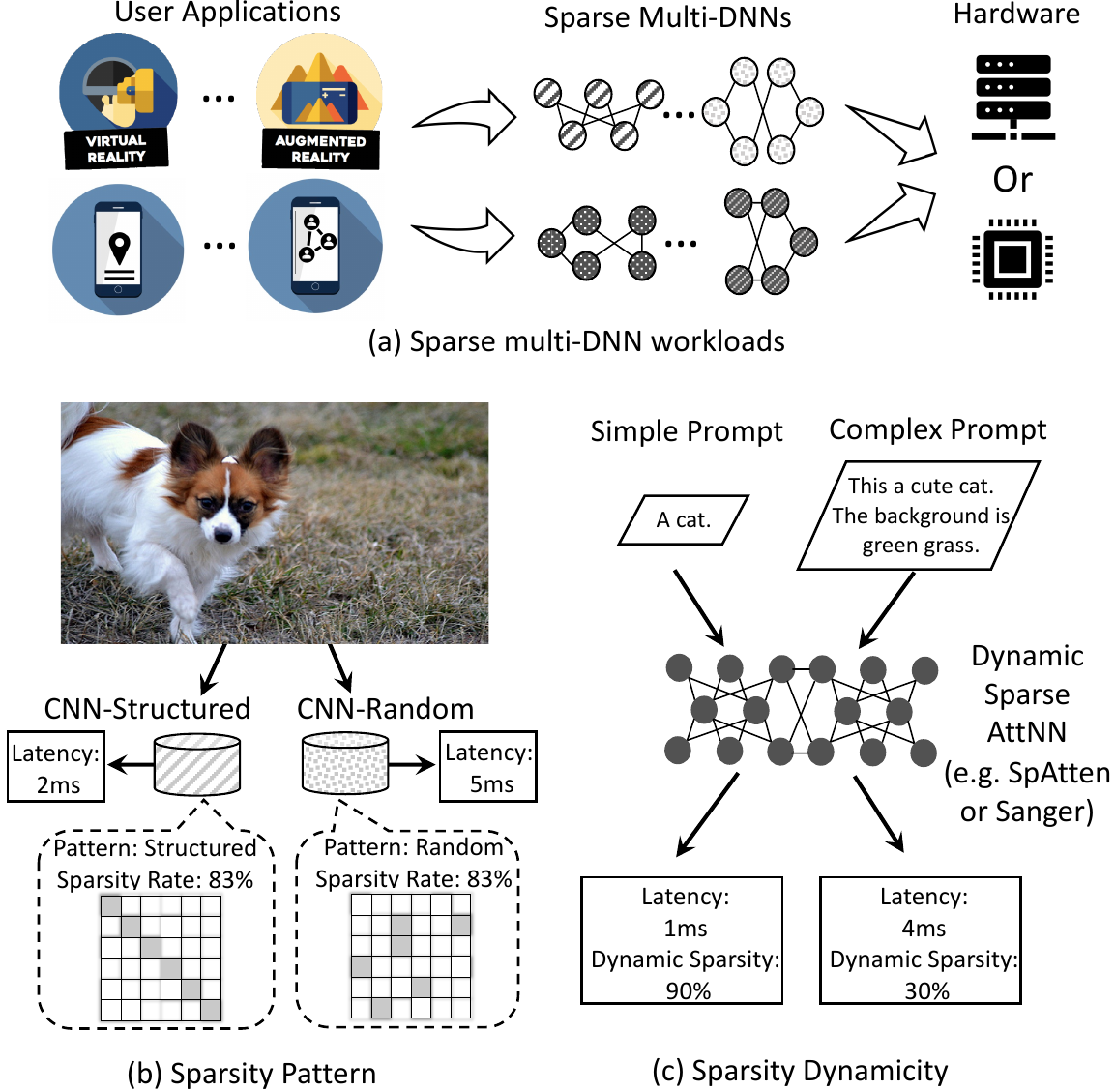}
\vspace{-4.0mm}
\caption{Sparse multi-DNN workloads with different sparsity patterns and runtime dynamicity.
}\label{fig:sparsity_pattern_dynamicity}
\vspace{-1mm}
\end{figure}

The unprecedented inference accuracy of DNNs has led to the rapidly increasing co-location of multiple DNN-powered applications~\cite{venieris2022multi}.
Spanning from mobile~\cite{smart_at_what_cost2021imc,band2022mobisys} and wearable apps~\cite{heimdall2020mobicom,xrbench2023mlsys} to cloud services~\cite{infaas2021atc,gpulet2022atc}, this trend leads to a significant rise in multi-DNN workloads.
Simultaneously, recent advances in model compression have enabled faster and smaller models through various sparsification techniques~\cite{menghani2023efficient, wang2019deep}. With inference workloads being commonly sparse, optimization opportunities have been explored, focusing on single-DNN execution~\cite{sparse_hw_survey2021jproc}.
This emergence of sparsity has also influenced AI hardware,
as evidenced by the sparsity support of NVIDIA GPUs via Tensor Core~\cite{nvidia2021accelerating} and Google TPU v4 via SparseCore~\cite{jouppi2023tpu}.
Nonetheless, despite its ubiquitousness across DNNs, sparsity has remained largely unexplored in multi-DNN workloads. 
As depicted in~\figref{fig:sparsity_pattern_dynamicity}(a),
the increasing number of parallel requests raised by diverse user applications has resulted in a surging demand for efficiently handling multiple sparse DNNs.
Therefore, there is an urgent need to study sparse multi-DNN workloads.

One of the key components while executing such multi-tenant workloads is the \textit{scheduler}.
Being responsible for deciding which task to dispatch next to the processing engine, extensive evaluation~\cite{choi2020prema,aimt2020isca,layerweaver2021hpca} has demonstrated that the quality of scheduling determines to a large degree the attainable performance under DNN multi-tenancy, rendering the design of the scheduler a critical task.
Although various multi-DNN scheduling approaches have been proposed~\cite{fcnnx2018fpl,aimt2020isca,ghodrati2020planaria,layerweaver2021hpca,kwon2021heterogeneous,kim2022sdrm3},
they are currently bounded by two main limitations:
\begin{itemize}[leftmargin=*]
    \item These scheduling approaches are optimized to perform well primarily on a single metric, \textit{e.g.}~either minimizing the average normalized turnaround time (ANTT)~\cite{eyerman2008system} or the latency service-level objective (SLO) violations, while excessively degrading the other (\secref{subsubsec:opt_opportunity}).
    \item The sparsity is largely neglected in these previous approaches, leading to suboptimal results. More specifically, these methods do not consider fine-grained details, such as sparsity patterns and dynamicity, that are crucial for further optimizations under sparse multi-DNN workloads.
\end{itemize}
In this paper, we aim to take the first step in systematically investigating the optimization opportunities in sparse multi-DNN workloads.
To achieve this, we propose a public sparse multi-DNN workload benchmark,
which includes convolutional neural networks (CNNs) for vision tasks and attention-based neural networks (AttNNs) for natural language processing (NLP) applications.
These benchmark models are sparsified using dynamic and static approaches with different patterns in order to emulate real-world scenarios. 
As shown in~\figref{fig:sparsity_pattern_dynamicity},
we identify two sparsity properties that lead to significant inter-model (across models) and intra-model (across input samples) dynamicity at runtime:
\begin{itemize} [leftmargin=*]
    \item Sparsity pattern, which refers to the pattern of non-zero masks used while sparsifying weights. \figref{fig:sparsity_pattern_dynamicity}(a) presents an example of CNNs with the same sparsity rate but different sparsity patterns, yielding varied latencies due to the count of effective operations.
    \item Sparsity dynamicity, which comprises the input-dependent sparsity in activations that leads to latency variability across input samples.~\figref{fig:sparsity_pattern_dynamicity}(b) illustrates an NLP example, where short and simple prompts with less information may lead to higher dynamic sparsity and lower latency compared to long and complex prompts at runtime. This category of dynamic sparsity is usually generated by either dynamic pruning approaches~\cite{sanger201micro, elsa2021isca, dota2022asplos} or certain operations, such as rectified linear units (ReLUs)~\cite{nair2010rectified}, that regularly produce zero values.
\end{itemize}
These two sparsity properties lead to the dynamic behavior of sparse DNNs during execution, which affects the scheduling design while optimizing SLO violation rate and ANTT. We provide further profiling and statistical analysis in~\secref{subsec:motivation}.

Based on the identified sources of sparsity dynamicity and the proposed benchmark, this paper proposes \textit{\sys}, a novel bi-level dynamic and static scheduler that optimizes for both ANTT and violation rate.
At the first software-based level, \sys utilizes static information, such as sparsity patterns and average latency, to obtain the initial task priorities before the execution of each task.
At the second level,
we introduce a lightweight hardware scheduler that adaptively refines the scheduling by monitoring the runtime information, such as the sample-specific dynamic sparsity, for further optimization.
\tabref{tb:comp_prev_work} presents a comparison of existing multi-DNN schedulers. 
To foster further development of multi-DNN acceleration research,
we intend to open-source our code, including benchmarks, evaluation infrastructure and proposed approach, and submit our work for artifact evaluation.
Overall, this work makes the following contributions:
\begin{itemize}[leftmargin=*]
  \item A public benchmark of sparse multi-DNN workloads that contains a wide range of DNNs with various forms of sparsity, including both dynamic and static sparsity and diverse sparsity patterns (\secref{sec:method_benchmark}).
  \item A bi-level dynamic and static scheduling framework, named \textit{Dysta}, which utilizes both static and dynamic sparsity information to optimize the processing of sparse multi-DNN workloads (\secref{sec:scheduler}).
  \item A low-cost hardware design of the \textit{Dysta} scheduler and an efficient sparse latency predictor. Several hardware optimizations are proposed to reduce the resource consumption of the scheduler, making its hardware overhead negligible compared with the overall accelerator (\secref{sec:hw_opt}).
\end{itemize}
To assess the effectiveness of our method, we conducted a comprehensive evaluation on the new benchmark we introduced (\secref{sec:evaluation}). Our approach achieves significant improvements in both ANTT and violation rate over the state-of-the-art multi-DNN schedulers, demonstrating its advantages and the value of the new benchmark for evaluating schedulers on sparse multi-DNN workloads.

\begin{table}[t]
    \centering
    \vspace{-2mm}
    \caption{Comparison of existing multi-DNN schedulers.}
    \vspace{2.0mm}
    \label{tb:comp_prev_work}
    \setlength\tabcolsep{1pt} 
    \scalebox{0.8}{
    \begin{tabular}{@{}L{2.75 cm} | C{1.8cm} C{1.8cm} C{2.8cm} C{1.5cm}@{}}
        \toprule
        {\textbf{Design}} & {\textbf{ANTT Optimized}} & {\textbf{SLO~Viol. Optimized}}& {\textbf{Dynamic \& Static Sparsity}}& {\textbf{Pattern Aware}} \\
        \midrule
        {PREMA}~{\cite{choi2020prema}} & {\ding{52}}& {\ding{55}}& {\ding{55}} &  {\ding{55}}\\
        {AI-MT}~{\cite{aimt2020isca}} & {\ding{52}}& {\ding{55}}& {\ding{55}}  & {\ding{55}}\\
        {Layerweaver}~{\cite{layerweaver2021hpca}} & {\ding{52}}& {\ding{55}}& {\ding{55}}  & {\ding{55}}\\
        {Planaria}~{\cite{ghodrati2020planaria}} & \ding{55}& \ding{52}& \ding{55}  & \ding{55}\\
        {SDRM\textsuperscript{3}}~{\cite{kim2022sdrm3}} & \ding{55}& \ding{52}& \ding{55} & \ding{55} \\
         \sys (Our Work) & \ding{52} & \ding{52}& \ding{52}& \ding{52}\\
        
        \bottomrule
    \end{tabular}}
    \vspace{-2.0mm}
\end{table}
\section{Background and Motivation}\label{sec:backgroud}

\subsection{Multi-DNN Workloads and Scheduling}
\label{subsec:backgrnd_multidnn_sched}

The scheduling strategy constitutes a core component of multi-DNN accelerators that directly affects the attainable performance of the system~\cite{venieris2022multi}. In the context of multi-DNN workloads with the layer-wise processing manner, the scheduler determines which layer of which model should be processed next and is the main driver of preemption and resource allocation decisions. Depending on the characteristics of the target application, scheduling can be either static or dynamic. 
Static scheduling~\cite{fcnnx2018fpl,kwon2021heterogeneous, magma2022hpca} is suitable for fixed-purpose multi-DNN systems, such as robots and autonomous vehicles, where the target set of DNNs and their performance requirements are known \textit{a priori}.
In contrast, dynamic scheduling~\cite{choi2020prema,ghodrati2020planaria,aimt2020isca,layerweaver2021hpca,kim2022sdrm3} is a more flexible approach, where dynamic schedulers allow new inference tasks to be processed and resources to be repurposed based on the completed and already running DNNs.

Table~\ref{tb:comp_prev_work} presents the existing multi-DNN dynamic scheduling algorithms~\cite{choi2020prema,ghodrati2020planaria,aimt2020isca,layerweaver2021hpca,kim2022sdrm3}. A key commonality behind them is their strong reliance on the assumption that DNN inference constitutes a predictable workload, \textit{i.e.}~it comprises a static computation graph for all inputs. Based on this assumption, execution time estimates, obtained through an offline profiling stage, are then employed to guide runtime scheduling decisions. Nonetheless, the static-workload assumption leads to overly conservative schedules and, in turn, to inefficient utilization of the hardware accelerator. As described in \secref{subsec:motivation}, sparsity constitutes an increasingly ubiquitous form of dynamicity in modern and upcoming DNNs that shifts the status-quo fixed DNN workload towards a dynamic, input-dependent workload, with the sparsity level changing in a per-sample manner. With multi-DNN workloads putting unprecedented pressure on computational demands, there is an emerging need for novel sparsity-aware solutions that extract the maximum performance from the underlying accelerator and enable the scalable processing of multiple DNNs.

\subsection{Sparse DNN Accelerators}\label{subsec:sparse_accel}

Stemming from sparsity-inducing training~\cite{sparse_induc_training2020icml}, activation functions~\cite{nair2010rectified} or pruning methods~\cite{state_of_pruning2020mlsys}, sparsity is widely observed across different types of models. 
At the hardware front, a series of works have proposed sparse DNN accelerators~\cite{sigma2020hpca,wang2020spatten,sanger201micro,gospa2021isca,jang2021sparsity,flexagon2023asplos} that manage to obtain speedup and efficiency gains through zero-skipping mechanisms and efficient sparse-storage schemes. 
Besides the static weight sparsity that is known \textit{a priori}, advanced sparse accelerators~\cite{sigma2020hpca,sanger201micro,gospa2021isca,jang2021sparsity,flexagon2023asplos} also support dynamically skipping ineffectual computations based on the input-dependent sparsity of activations, providing additional performance improvements over conventional accelerators. 

\subsection{Motivation}\label{subsec:motivation}

In the context of multi-DNN workloads, previous literature has so far overlooked
the various types of sparsity that are observed across models.
In this paper, we identify two primary properties of sparsity, namely \textit{1)}~sparsity dynamicity and \textit{2)}~sparsity pattern, and study their effects in scheduling sparse multi-DNN workloads.

\begin{figure}[t]\centering
    \includegraphics[width=0.49\textwidth]{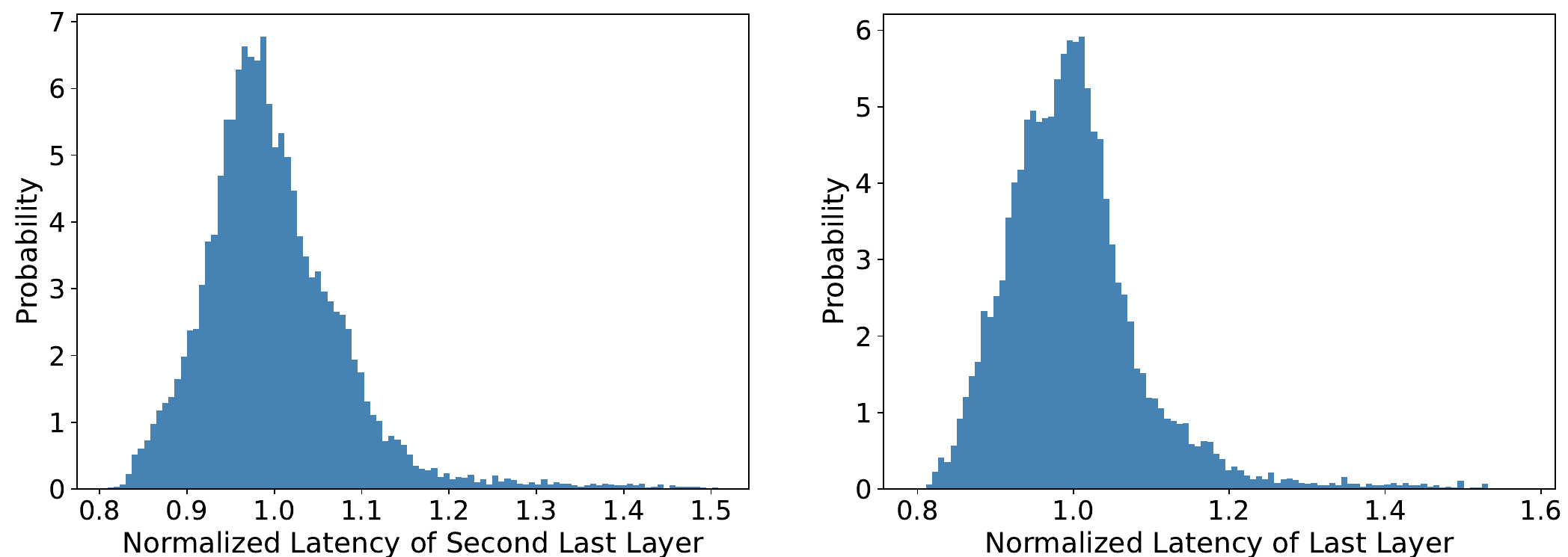}
    \vspace{-4.0mm}
    \caption{Impact of dynamic sparsity on language models.}\label{fig:sparsity_bert_sanger}
    \vspace{-2.0mm}
\end{figure} 

\subsubsection{Sparsity Dynamicity }\label{subsubsec:dyn_sparsity}

Sparsity dynamicity is widely observed across different families of DNNs, including both attention-based (AttNNs) and convolutional neural networks (CNNs).
The two main sources of dynamicity are \textit{i)}~dynamic pruning methods and \textit{ii)}~operations that regularly generate zero values.

For AttNNs, the most common form of sparsity dynamicity comes from dynamic pruning approaches.
By focusing on weak connections between different tokens in the attention matrix,
various pruning techniques have been proposed to exploit the attention sparsity for acceleration~\cite{ham2020, wang2020spatten, sanger201micro, zhou2021energon, elsa2021isca, dota2022asplos}.
To demonstrate the dynamic behavior introduced by the attention sparsity,
we profile the per-layer latency of  \textit{BERT}~\cite{devlin2018bert} on the accelerator introduced by~\cite{sanger201micro}.
We iterate the sparse \textit{BERT} over the SQUAD~\cite{rajpurkar2016squad} dataset.
For better visualization,
we normalize the distribution by the average latency.
As it can be seen in~\figref{fig:sparsity_bert_sanger},
the normalized latency varies from $0.6$ to $1.8$, resulting in significant dynamic behavior during runtime.

\begin{figure}[t]\centering
    \includegraphics[width=0.43\textwidth]{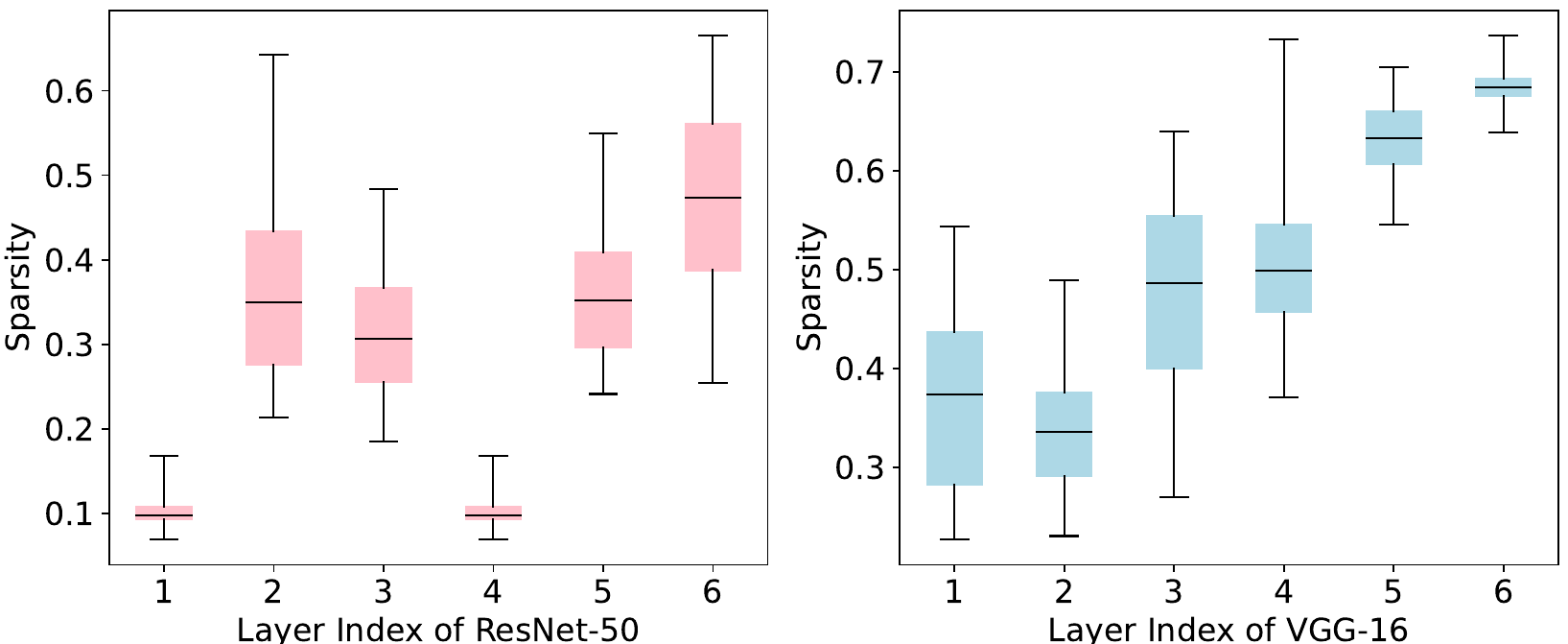}
    \vspace{-1.0mm}
    \caption{Sparsity ratios of \textit{ResNet-50} and \textit{VGG-16}.}\label{fig:sparsity_layer_relu}
    \vspace{-1.0mm}
\end{figure} 

\begin{table}[t]
    \centering
    \caption{Relative range of network sparsity. }
    \vspace{0.2cm}
    \label{tb:network_sparse}
    \setlength\tabcolsep{1pt}
    \scalebox{0.78}{
    \begin{tabular}{@{}C{2.8cm}|C{1.8cm}|C{1.8cm}|C{1.8cm}|C{1.8cm}@{}}
        \toprule
        \midrule
        \textbf{Model} & \textit{GoogLeNet} & \textit{VGG-16} & \textit{InceptionV3} & \textit{ResNet-50} \\ \midrule 
        \textbf{Relative Range} & 28.3\% & 21.8\% & 23.0\% & 15.1\% \\ 
        \bottomrule
    \end{tabular}
    }
\end{table}

For CNNs, 
we focus on the ReLU activation function~\cite{nair2010rectified} as an example of an operation that regularly generates zeros at runtime and study its dynamic behavior.
Although PREMA~\cite{choi2020prema} has briefly discussed the impact of ReLU on runtime dynamicity in CNNs,
it did not consider out-of-distribution or less informative inputs that may cause large sparsity variances, such as images taken by users in dark or poorly illuminated environments.
In this paper, we conduct a more comprehensive analysis by including low-light images from the ExDark~\cite{loh2019getting} and DarkFace~\cite{Chen2018Retinex} datasets to emulate real scenarios.
We profile the activation sparsity of the last six layers in \textit{ResNet-50} and \textit{VGG-16}.
As shown in~\figref{fig:sparsity_layer_relu},
the sparsity ratios of most layers range from $10$\% to $45$\%, indicating that a large variance is introduced when considering out-of-distribution or less informative images.
To investigate the sparsity of the network instead of individual layers,
we calculate the network sparsity by averaging individual layer sparsities across the whole network.
As shown in~\mbox{\tabref{tb:network_sparse}},
the relative range of network sparsity can reach up to $28.4$\%, depending on the model.

\begin{figure}[t]\centering
    \includegraphics[width=0.49\textwidth]{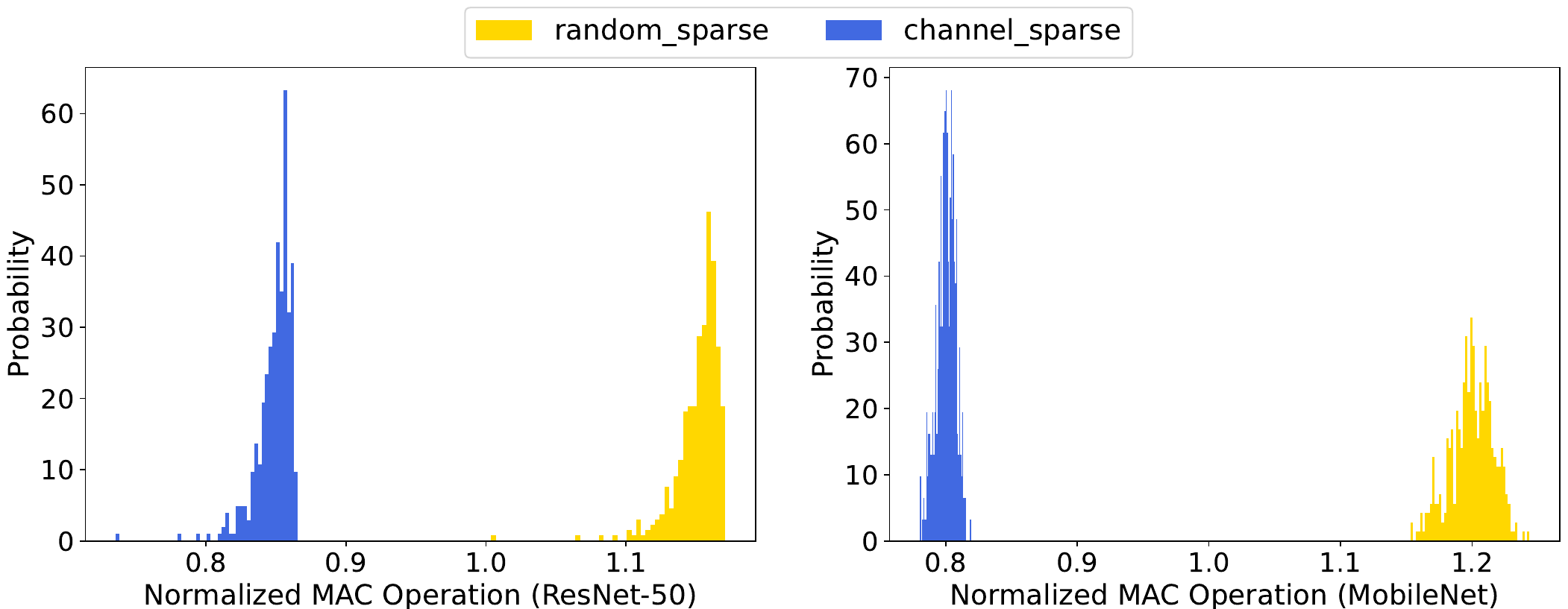}
    \vspace{-4.0mm}
    \caption{Impact of weight sparsity pattern on the valid MAC operations on \textit{ResNet-50} and \textit{MobileNet}.}\label{fig:sparsity_pattern}
    \vspace{1.0mm}
\end{figure}

\subsubsection{Sparsity Pattern}\label{subsubsec:sparsity_pattern}

Sparsity pattern refers to the mask type used when sparsifying DNNs.
Mainstream sparsity patterns include point-wise random~\cite{han2015deep}, block-wise structured~\cite{zhou2021learning, fan2022adaptable} and channel-wise sparsification~\cite{he2017channel}.  
The support for different sparsity patterns largely depends on the underlying hardware design.
For instance, 
the Sparse Tensor Core introduced in the NVIDIA Ampere architecture~\cite{nvidia2021accelerating} supports the block-wise N:M pattern, while other customized accelerators have better performance for the point-wise random pattern~\cite{chen2019eyeriss, wu2022sparseloop}.
Although previous work attempts to leverage sparsity information when scheduling multi-DNN workloads,
they only rely on the overall sparsity ratio, without considering the sparsity patterns and their relationship with the target hardware.

To investigate the effect of sparsity patterns,
we profile the amount of valid MAC operations introduced by point-wise random and channel-wise sparsities, respectively.
For both patterns, we keep the same overall sparsity ratio with identical input images for profiling.
\mbox{\figref{fig:sparsity_pattern}} presents the distribution of normalized MAC operations on \textit{ResNet-50} and \textit{MobileNet} with sparsity ratios of $95$\% and $80$\%, respectively.
Notably, different sparsity patterns may introduce up to $40$\% difference in normalized valid MACs, despite the same sparsity ratio.
\vspace{2mm}

\subsubsection{Optimization Opportunities}\label{subsubsec:opt_opportunity}

Given our analysis, we identify two key optimization opportunities for the scheduler of sparse multi-DNN workloads.

\begin{figure}[t]\centering
\includegraphics[width=0.49\textwidth]{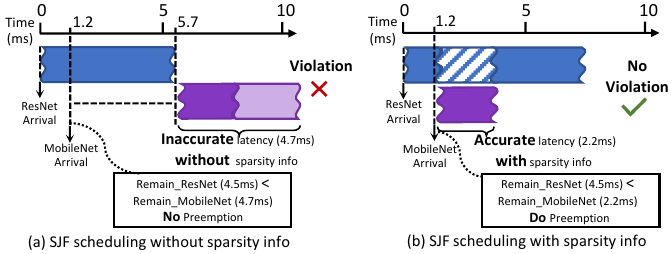}
\vspace{-3.0mm}
\caption{SJF schedulings with and without sparsity information in task violation.}\label{fig:sparse_schedule_example}
\end{figure}

\Hsection{Opportunity One: Dynamicity- \& Pattern-Aware}
Based on the profiling results of Sections~\ref{subsubsec:dyn_sparsity} and~\ref{subsubsec:sparsity_pattern},
sparsity may lead to significant dynamic behavior at runtime, which affects the optimality of schedulers when targeting sparse multi-DNN workloads.
\figref{fig:sparse_schedule_example} presents an example of the Shortest-Job First (SJF) scheduler, to illustrate the importance of capturing sparsity information for multi-DNN workloads.
When exploiting fine-grained information, such as dynamic sparsity ratio or sparsity pattern,
the scheduler can make a more informed preemption decision based on accurate latency estimation, avoiding the violation of the second request.
Based on this observation, in this work, we aim to utilize sparsity dynamicity and pattern information to improve sparse multi-DNN scheduling.

\Hsection{Opportunity Two: SLO- \& ANTT-Optimized}
Although previous approaches have attempted to optimize multi-DNN scheduling,
we observe that they only perform well on a single metric,
\textit{e.g.}~either the latency SLO violation rate or ANTT, while sacrificing the performance of the other.
For instance, compared with the traditional SJF scheduler,
current SOTA approaches can only improve ANTT under the condition of excessively higher SLO violations~\cite{choi2020prema}.
The reasons are that they either \textit{i)}~are not deadline-aware or \textit{ii)}~do not consider the optimization of the violation rate or ANTT while including deadlines.
In this paper,
we aim to address these drawbacks by co-optimizing both SLO violation rate and ANTT.

\section{Sparse Multi-DNN Benchmark}\label{sec:method_benchmark}

\subsection{Benchmark Models}\label{subsec:benchmark}
One of the limitations in previous multi-DNN research is the absence of publicly accessible benchmarks
for conducting fair comparisons.
To address this issue,
we propose a public benchmark for sparse multi-DNN scheduling.
Our benchmark is designed to accommodate a diverse collection of sparsified CNNs and AttNNs. 

As shown in~\tabref{tb:sparse_benchmark},
we consider sparse multi-DNN workloads in three setups: data center, mobile phones and AR/VR wearables.
Our benchmark comprises six tasks spanning three distinct applications: visual perception, personal assistant, and hand tracking.
The tasks include object detection, image classification, machine translation, question \& answering, hand tracking and gesture recognition.
For vision tasks, we adopt four popular CNN models, namely \textit{SSD}~\cite{liu2016ssd}, \textit{ResNet-50}~\cite{he2016deep}, \textit{VGG-16}~\cite{simonyan2015very}, and \textit{MobileNet}~\cite{mobilenet2017arxiv}.
For \mbox{AttNNs},
our benchmark includes three commonly used language models: \textit{BERT}~\cite{devlin2018bert}, \textit{BART}~\cite{lewis2019bart}, and \textit{GPT-2}~\cite{radford2019language}.
In terms of datasets,
we adopt ImageNet~\cite{deng2009imagenet}, ExDark~\cite{loh2019getting}, DarkFace~\cite{Chen2018Retinex}, and COCO~\cite{lin2014microsoft} for training and evaluation on vision tasks, while
GLUE~\cite{wang2018glue} and SQUAD~\cite{rajpurkar2016squad} are used for language tasks.

\begin{table}[t]
    \centering
    \vspace{-2mm}
    \caption{Benchmark Models. }
    \vspace{0.2cm}
    \label{tb:sparse_benchmark}
    \setlength\tabcolsep{1pt}
    \scalebox{0.78}{
    \begin{tabular}{@{} L{2.2cm} L{3.0cm} L{3.4cm} L{2.5cm} @{}}
        \toprule
        \midrule
        \textbf{Scenarios} &  \textbf{Applications} &  \textbf{Tasks}  & \textbf{Models} \\ 
        \midrule
         
         \multirow{2}{*}{{Data Center}}& \multirow{2}{*}{Visual Perception} & Object Detection & \textit{SSD} \\ 
         \cmidrule{3-4}
         &  & Image Classification & \textit{VGG-16, ResNet-50} \\  
         \midrule
        \multirow{2}{*}{{Mobile Phone}}& \multirow{2}{*}{Personal Assistant}  & Machine Translation & \textit{BART, GPT-2} \\ 
        \cmidrule{3-4}
        & & Question~\&~Answering & \textit{BERT} \\ 
        \midrule 
        {{AR/VR}} & \multirow{2}{*}{Hand Tracking} & Hand Detection & \textit{SSD} \\ 
        \cmidrule{3-4}
        {{Wearables}} &  & Gesture Recognition & \textit{MobileNet} \\  
        \bottomrule
    \end{tabular}
    }
\end{table}

\subsection{Model Sparsification}\label{subsec:model_sparsification}
\begin{figure}[htp]
    \centering
    \includegraphics[width=0.49\textwidth]{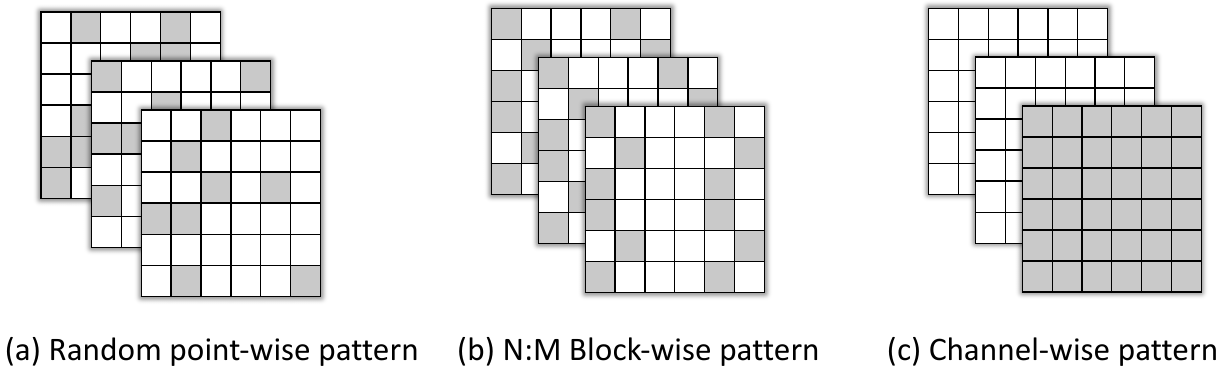}
    \vspace{-3.0mm}
    \caption{Three sparsity patterns applied on CNNs.}\label{fig:sparse_patterns}
    \vspace{2.0mm}
\end{figure}

In order to study the impact of sparsity patterns,
we adopt three different pruning methods for CNNs:
random point-wise~\cite{han2015deep}, N:M block-wise~\cite{zhou2021learning} and channel-wise pruning~\cite{he2017channel}.
The generated sparsity patterns of each pruning approach are shown in~\figref{fig:sparse_patterns}.
We obtain the pre-trained CNNs from \textit{PyTorch}~\cite{pytorch},
and apply the target pruning methods using sparsification recipes provided by \textit{SparseML}~\cite{kurtz2020inducing} and \textit{SparseZoo}.\footnote{\url{https://sparsezoo.neuralmagic.com/}}
We expose the sparsity rate as a tunable parameter.

For AttNNs,
we adopt the dynamic pruning approach introduced in~\cite{sanger201micro} to study the effect of sparsity dynamicity. 
The adopted method performs pruning dynamically via binary thresholding based on a lightweight prediction of the attention
matrix.
Following their open-source code,\footnote{\label{sanger_link}\url{https://github.com/hatsu3/Sanger}}
we set the threshold as $0.2$ for \textit{BART}, and $0.002$ for \textit{BERT} and \textit{GPT2} to maintain the original accuracy of each model.

\subsection{Evaluation Methodology}\label{subsec:evaluate_method}

\subsubsection{Methodology Overview}

\begin{figure}[t]\centering
    \includegraphics[width=0.46\textwidth]{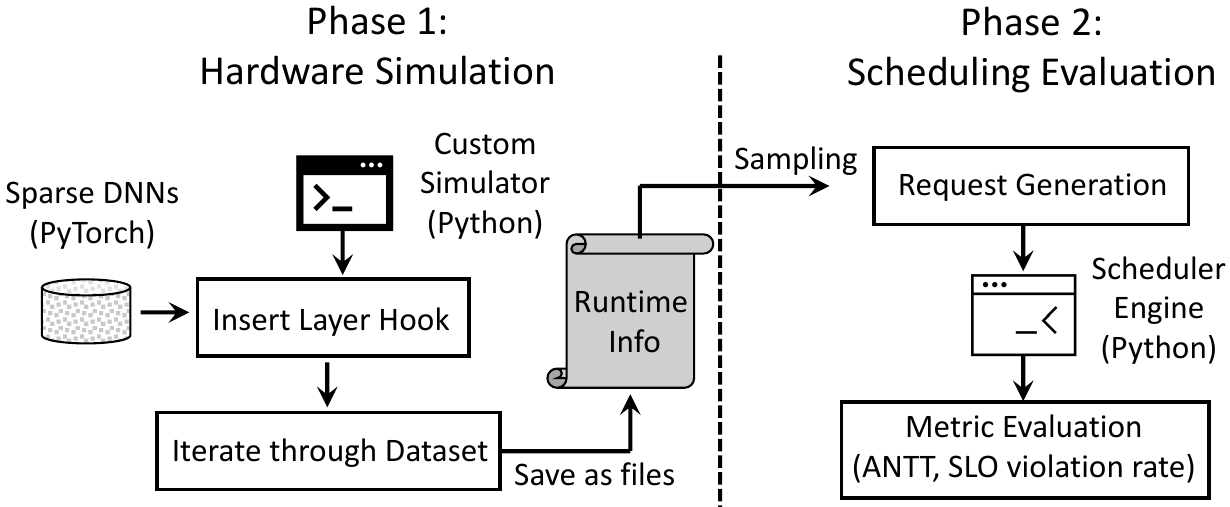}
    \caption{Evaluation methodology.}\label{fig:overview_evaluation}
    \vspace{2.0mm}
\end{figure}

\figref{fig:overview_evaluation} shows our evaluation process, comprising hardware simulation and scheduling evaluation phases.
During the \textit{Hardware Simulation} phase,
we insert our \textit{Python}-based hardware simulator into sparse models using the \textit{Hook} function provided by \textit{PyTorch}.
This enables the generation of runtime information such as 
per-layer latency and sparsity ratio of each model for the target sparse hardware.
We obtain the runtime information for each input and model pair by processing the whole dataset with each sparse model.
The runtime information is then saved as files for later use.
In the \textit{Scheduling Evaluation} phase,
we generate multiple task requests by sampling from sparse models summarized in Sections~\ref{subsec:benchmark} and~\ref{subsec:model_sparsification}.
The \textit{Scheduler Engine} is then deployed to handle incoming requests according to the designated scheduling algorithm.
The runtime information obtained in the hardware simulation phase is used by the \textit{Scheduler Engine} to simulate the execution of the target hardware.
After the completion of all jobs, we evaluate different metrics, including both ANTT and violation rate based on the user-specified latency SLO.

\subsubsection{Hardware Simulator}
To simulate sparse AttNNs,
we select \textit{Sanger}~\cite{sanger201micro} as the target hardware accelerator with the support of load-balanced computation for attention sparsity.
Their open-source simulator\textsuperscript{\ref{sanger_link}} is adopted in our evaluation.
For CNNs, although there are various sparse accelerators,
only a few of them open-source their design or simulator.
The most promising third-party simulators are \textit{SparseLoop}~\cite{wu2022sparseloop} and \textit{STONNE}~\cite{munoz2021stonne}.
However, these frameworks suffer from either prohibitively slow evaluation when actual data are used or limited \textit{PyTorch} support,
rendering them unsuitable for meeting our needs.
As a result, we develop a custom \textit{Python}-based simulator for sparse CNN evaluation.
In this endeavor, we select \textit{Eyeriss-V2} as the target CNN accelerator for two reasons.
First, it supports both weight and activation sparsity. 
Second, several third-party implementations are available for replicating the \textit{Eyeriss-V2} hardware design.\footnote{\url{https://github.com/SingularityKChen/dl_accelerator} and \url{https://github.com/karthisugumar/CSE240D-Hierarchical_Mesh_NoC-Eyeriss_v2}}
This allowed us to use these references to validate the correctness of our custom simulator.

\begin{figure}[t]\centering
    \includegraphics[width=0.49\textwidth]{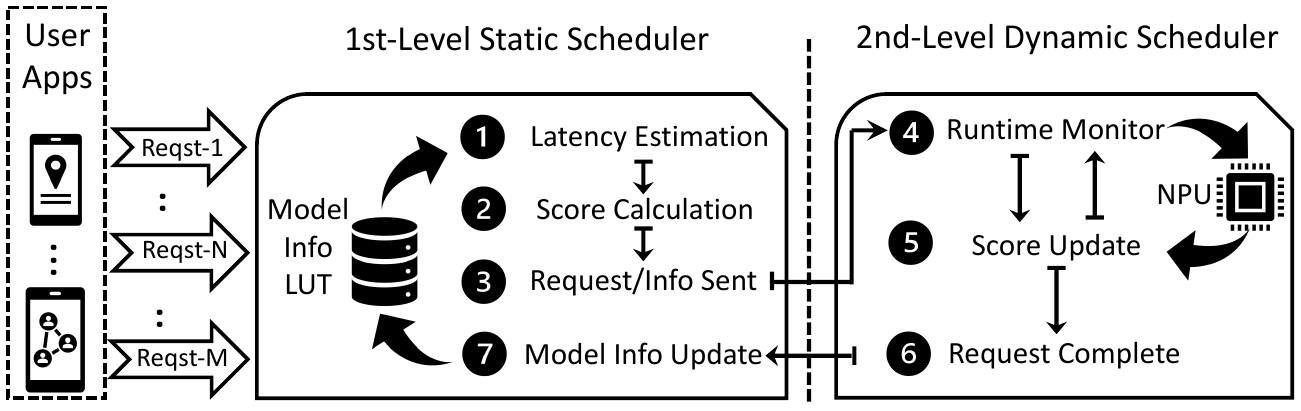}
    \vspace{-4.0mm}
    \caption{Overview of \sys's scheduling approach.}\label{fig:overview_scheduling}
    \vspace{-2.0mm}
\end{figure} 

\section{Dynamic and Static Scheduler}\label{sec:scheduler}

\subsection{Framework Overview}\label{subsec:overview}

As demonstrated in~\secref{subsubsec:opt_opportunity}, the dynamicity and pattern of sparsity introduce variations in runtime behavior that can significantly impact the optimality of scheduling.
To alleviate this,
we introduce \sys, a bi-level scheduling method that combines static and dynamic scheduling while leveraging information about both sparsity dynamicity and pattern.

\figref{fig:overview_scheduling} presents an overview of the proposed method.
Our bi-level scheduler consists of a static and a dynamic component, implemented at the software and hardware levels, respectively.
At the first level, the static component assigns the incoming requests with initial scores to determine the processing order, taking into account the sparsity patterns of the DNN workloads.
At this step, the static scheduler also populates a lookup table (LUT) with each request's model information. This per-request information includes \textit{i)}~the model's sparsity pattern, \textit{ii)}~the average sparsity ratio across all layers, and \textit{iii)}~the average latency on the target hardware.
The average sparsity and latency information is obtained by profiling representative requests offline.
At the second level, the dynamic component interacts with the target hardware accelerator (neural processing unit (NPU) in \figref{fig:overview_scheduling}), to monitor the input-dependent, runtime sparsity of the workload.
The obtained information is then used to update the task scores and adjust their processing order dynamically.
We delve deeper into the details of \sys's static and dynamic components in the next subsection.

\subsection{Dysta Scheduling Algorithm}\label{subsec:dysta_scheduler}

\subsubsection{Static Scheduler}\label{subsubsec:static_scheduler}

Algorithm~\ref{alg:dysta_static_sched} describes the proposed software-based static scheduler.
Upon the arrival of a new request $Reqst_n$, defined as a tuple $\left<Model_n, Pattn_{n}, input_n, SLO_n\right>$, 
the static scheduler assigns an initial score $Score_n$ that determines its processing order before knowing any runtime information. 
To meet the multi-objective requirements of multi-DNN systems and maintain a balance between ANTT and SLO violation rate, we express the initial score of the $n$-th request as the weighted sum of two factors (line~7): the estimated latency $Lat_n$ (line~5), and the slack time $T_n^{\text{Slack}}$ that is determined based on the request's latency SLO (line~6).

Each of the two factors serves a different purpose. First, by incorporating the estimated latency in the score, we encourage shorter jobs to finish earlier. As such, it yields schedules that primarily improve the ANTT metric. Second, by integrating the slack into the score, our scheduler prioritizes jobs with tight deadlines to be dispatched sooner. Thus, the latency SLO violations are kept to a minimum.
Overall, to balance the optimization between ANTT and SLO violation rate, we parametrize our formulation by means of hyperparameter $\beta$ (line~7), which allows us to tunably weight the two factors.

Moreover, to capture the effect of sparsity pattern,
latency is estimated separately for each model-pattern pair.
We use the average latency to estimate $Lat_n$, which is obtained by collecting latency information either from actual executed requests or from using the performance model on a number of representative requests.
A LUT is used to store the average latency and sparsity pattern pair of each model. As such, the estimated latency can be quickly obtained by accessing the LUT's entry for the given model-pattern pair.
Finally, the calculated score and the request information are passed to the hardware-based dynamic scheduler for execution.

\SetArgSty{textnormal}
\begin{algorithm}[!t]
    \setcounter{AlgoLine}{0}
    \scriptsize
    \SetAlgoLined
    \LinesNumbered
    \DontPrintSemicolon

    \textbf{Input:} $Reqst_{n} = \left<Model_n, Pattn_{n}, input_n, SLO_n\right>$

    \textbf{Output:} $Score_{n}, Info_{n}$ 
    
    \If(\Comment*[f]{\scriptsize \textrm{New request arrives}}){$Reqst_{n}$ arrives}{
        $Info_n \leftarrow \text{LUT}(Reqst_{n})$ \Comment*[f]{\scriptsize \textrm{Obtain model info from LUT}}
    
        $Lat_n \leftarrow \text{PredLat}(Model_{n}, Pattn_{n})$ \Comment*[f]{\scriptsize \textrm{Estimate latency w/ sparsity pattern}}
        
        $T_{n}^{\text{Slack}} \leftarrow SLO_{n} - Lat_{n}$ \Comment*[f]{\scriptsize \textrm{Calculate slack time}}

       $Score_{n} \leftarrow Lat_{n} + \beta \times T_{n}^{\text{Slack}}$ \Comment*[f]{\scriptsize \textrm{Calculate score}}

       $\text{SendHW}(Score_{n}, Info_{n})$ \Comment*[f]{\scriptsize \textrm{Forward score \& info to dynamic scheduler}}
    }
    \caption{Static Scheduler}
    \label{alg:dysta_static_sched}
\end{algorithm}

\subsubsection{Dynamic Scheduler}\label{subsubsec:dynamic_scheduler}

Algorithm~\ref{alg:dysta_dynamic_sched} presents the hardware-based dynamic scheduler.
The dynamic scheduler uses a queue $ReqstQ$ to store all the incoming requests forwarded by the static scheduler.
Our scheduling assumes that the execution is performed in a per-layer or per-layer-block manner, which is a common setting in commercial AI hardware~\cite{nvidia2021accelerating, jouppi2023tpu} and existing research literature~\cite{chen2016eyeriss,chen2019eyeriss,sanger201micro,dota2022asplos,fan2022adaptable}.
Thus, whenever the execution of one layer or layer block completes,
the dynamic scheduler is invoked to update the estimated latency and determine the next running request.
As shown on line~7, the updated latency estimate, $\widehat{T}^{\text{Remain}}_{cur}$, is obtained based on the monitored sparsity information $S^{{\text{Monitor}}}_{\text{cur}}$ using an efficient sparse latency predictor, PredSparseLat$(\cdot)$, detailed in~\secref{subsec:sparse_lat_pred}.
To determine the next request to process, the dynamic scheduler updates the score of each $Reqst_i$ from $ReqstQ$, and the next running request is selected as the one with the lowest score.

The score calculated by the dynamic scheduler comprises three terms (line~9): \textit{1)}~remaining time $\widehat{T}^{\text{Remain}}_i$, \textit{2)}~slack time ${T}^{\text{Slack}}_i$, and \textit{3)}~penalty ${T}^{\text{Penalty}}_i$.
The first two terms aim to improve ANTT and violation rate, respectively, similarly to the static scheduler. 
However, a key characteristic of the dynamic component is that it estimates latency $\widehat{T}^{\text{Remain}}_i$ with higher accuracy by relying on the actual monitored sparsity information and the sparse latency predictor. 
This approach allows us to alleviate the limitations demonstrated in \secref{subsubsec:opt_opportunity}, leading to better-informed scheduling decisions.
To avoid excessively frequent preemptions, we introduce the penalty term (line~10). We define this term as the ratio between the waiting time ${T}^{\text{Wait}}_i$ and the isolated execution time ${T}^{\text{Isol.}}_i$, normalized by the number of requests in queue $ReqstQ$.
A lower penalty indicates a request waiting for a shorter time, which encourages the scheduler to keep the nearest-executed request executing.
A hyperparameter $\eta$ is used to balance these terms in the score, allowing for a tunable trade-off between ANTT and SLO violation rate.

\SetArgSty{textnormal}
\begin{algorithm}[!t]
    \setcounter{AlgoLine}{0}
    \scriptsize
    \SetAlgoLined
    \LinesNumbered
    \DontPrintSemicolon

    
    \textbf{Input:} $Reqst_{n} = \left<Score_{n}, Info_{n}\right>$ \Comment*[f]{\scriptsize \textrm{Input request from static scheduler}}

    \textbf{Output:} $Reqst_{\text{next}}$ 

    \If(\Comment*[f]{\scriptsize \textrm{Request received, parallel execute with line~6}}){$Reqst_{n}$ received}{       

       $\text{Push}(ReqstQ, Score_{n}, Info_{n})$ \Comment*[f]{\scriptsize \textrm{Push request into queue}}
    }

    \If(\Comment*[f]{\scriptsize \textrm{Invoke scheduler}}){LayerRun($Reqst_{\text{cur}}$) return at time $t$}{

        $\widehat{T}^{\text{Remain}}_{\text{cur}} \leftarrow \text{PredSparseLat}(S^{{\text{Monitor}}}_{\text{cur}})$ \Comment*[f]{\scriptsize \textrm{Update remain time}}
        
        \ForEach(\Comment*[f]{\scriptsize \textrm{Update per-request score}}){$Reqst_{i}$ \textbf{in} $ReqstQ$} {

            
            $T^{\text{Slack}}_{i} \leftarrow SLO_{i} - t - \widehat{T}^{\text{Remain}}_i$

            $T^{\text{Penalty}}_{i} \leftarrow \left(T^{\text{Wait}}_i / T^{\text{Isol.}}_i\right) / |ReqstQ|$


            $Score_i \leftarrow \widehat{T}^{\text{Remain}}_i + \eta \times ( T^{\text{Slack}}_{i} + T^{\text{Penalty}}_{i})$
        }
        $Reqst_{\text{next}} \leftarrow \argmin\limits_{i}~ Score_i, \quad \forall i \in \bigl[1, |ReqstQ|\bigr]$ \Comment*[f]{\scriptsize \textrm{Select new request}}
        
    }
    \caption{Dynamic Scheduler}
    \label{alg:dysta_dynamic_sched}
\end{algorithm}

\section{Hardware Design}\label{sec:hw_opt}

In this section, we focus on the hardware design of \textit{Dysta}'s dynamic scheduler. We start by presenting the design of our sparse latency estimator, which constitutes a core driver behind the scheduler's decisions. Then, we proceed with the description of the microarchitecture, key components and hardware optimizations of our scheduler's design. 

\subsection{Sparse Latency Predictor}\label{subsec:sparse_lat_pred}

Designing the latency predictor for the sparse multi-DNN scheduler requires consideration of two metrics: accuracy and hardware overhead.
As illustrated in~\secref{subsubsec:opt_opportunity},
accurate latency estimation is crucial for improving the scheduling of sparse multi-DNN workloads.
However, we must also take into account the hardware overhead of the latency predictor.
To this end, although several advanced learning-based approaches, such as Gaussian Processes~\cite{gp_perf_model2021electronics}, Random Forests~\cite{nnmeter2021mobisys} and DNNs~\cite{brpnas2020neurips}, can be used for this task, the overhead of these methods is prohibitively costly for our scheduler operating at the layer granularity.

\begin{figure}[t]\centering
    \vspace{-3mm}
    \includegraphics[width=0.49\textwidth]{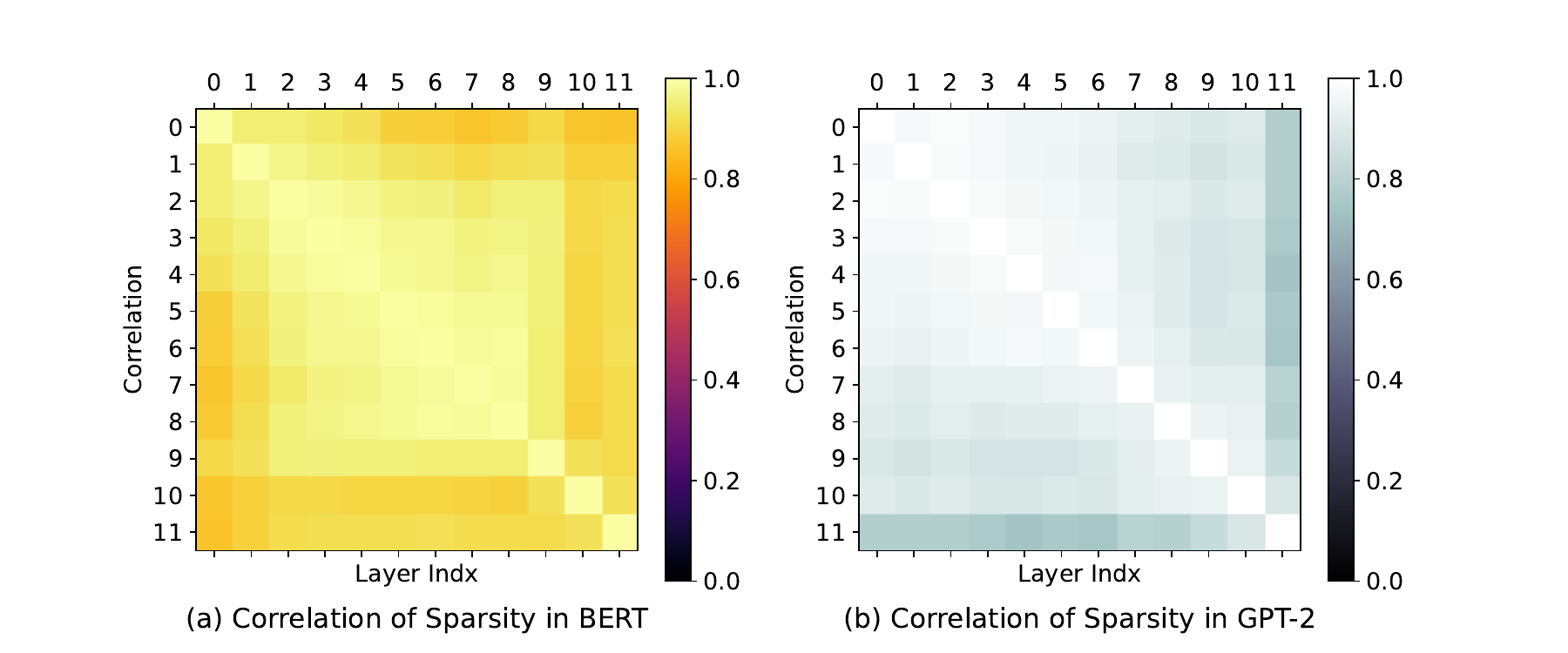}
    \vspace{-4.0mm}
    \caption{Correlation of sparsity of different layers.}\label{fig:corr}
    \vspace{-2.0mm}
\end{figure}

To design an accurate and efficient latency predictor for the sparse multi-DNN scheduler,
two key problems need to be addressed: 
\textit{1)}~identifying what sparsity information needs to be captured during runtime, and \textit{2)}~determining the best algorithm for latency prediction.
To this end,
we profile the layer sparsity of two popular AttNN models, \textit{BERT} and \textit{GPT-2} on SQUAD~\cite{rajpurkar2016squad} and GLUE~\cite{wang2018glue} datasets, respectively, and analyse their Pearson product-moment correlation in~\figref{fig:corr}.
The results indicate that the sparsities of different layers are highly linearly correlated in both models.
This observation motivates us to monitor the layer sparsity at runtime and adopt a linear model for sparse latency prediction.

\SetArgSty{textnormal}
\begin{algorithm}[t]
    \setcounter{AlgoLine}{0}
    \scriptsize
    \SetAlgoLined
    \LinesNumbered
    \DontPrintSemicolon

    \textbf{Input:} $S^{\text{Monitor}}, i, j$ \Comment*[f]{\scriptsize \textrm{Monitored information from NPU}}

    \textbf{Output:} $Lat_{i}^{\text{Sparse}}$ 


    \If(\Comment*[f]{\scriptsize \textrm{Monitored layer sparsity}}){$S^{\text{Monitor}}$ captured for $j$-th Layer of $Reqst_{i}$}{
        $S^{{\text{Avg}}}_{(i, j)} \leftarrow \text{SparsityLUT}(i, j)$ \Comment*[f]{\scriptsize \textrm{Get average sparsity}}

        $Lat^{\text{Avg}}_{i} \leftarrow \text{LatencyLUT}(i)$ \Comment*[f]{\scriptsize \textrm{Get average latency}}    

        $\gamma_{i} \leftarrow \text{PredSparsityCoeff}(S^{\text{Monitor}}, S^{\text{Avg}}_{(i, j)})$ \Comment*[f]{\scriptsize \textrm{Update sparsity coefficient}}
        
        $Lat_{i}^{\text{Sparse}} \leftarrow \alpha \times \gamma_{i} \times Lat_{i}^{\text{Avg}}$ \Comment*[f]{\scriptsize \textrm{Predict latency}}
    
    }
    \caption{Sparse Latency Prediction}
    \label{alg:sparse_pred}
\end{algorithm}

Algorithm~\ref{alg:sparse_pred} outlines our approach.
We deploy a hardware monitor to calculate the layer sparsity $S^{\text{Monitor}}$ for the current running  model (line~1).
The sparse latency predictor retrieves the average layer sparsity $S^{\text{Avg}}_{(i,j)}$ and latency $Lat^{\text{Avg}}_{i}$ from the sparsity and latency LUTs, respectively, based on the model-pattern pair information (lines~4 \& 5).
The average latency stored in the latency LUT is received from the static scheduler,
while the sparsity LUT is constructed by either collecting information from executed requests or obtaining information from model providers.
Whenever the hardware monitor returns the layer sparsity,
the sparsity coefficient $\gamma_i$ is calculated (line~6), representing the linear rate between monitored and average layer sparsities. 
We then estimate the latency as $\alpha \times \gamma_{i} \times Lat_{i}^{\text{Avg}}$,
where parameter $\alpha$ reflects how effectively sparsity can deliver real latency reduction.
The value of $\alpha$ depends on the underlying hardware and needs to be set per pattern.
As the accelerators targeted in this paper support both activation and weight sparsity,
we set $\alpha=1$.

\begin{table}[t]
    \centering
    \vspace{-2mm}
    \caption{Root-Mean-Square Error (RMSE) of sparse latency predictor using three different strategies. }
    \vspace{0.2cm}
    \label{tb:rmse}
    \setlength\tabcolsep{1pt}
    \scalebox{0.85}{
        \begin{tabular}{@{} L{2.3cm} L{2.3cm} L{2.3cm} L{2.3cm} @{}}
             \toprule
             & \textbf{\textit{Average-All}} & \textbf{\textit{Last-N}} & \textbf{\textit{Last-One}} \\ 
             \midrule 
             \textbf{BERT} & 0.000286 & 0.000419 & 0.000252 \\ 
             \textbf{GPT-2} & 0.000218 & 0.000421 & 0.000226 \\ 
             \bottomrule
        \end{tabular}
    }
\end{table}

To determine the sparsity coefficient $\gamma$, we consider three different approaches, namely \textit{average-all}, \textit{last-N}, and \textit{last-one}.
The \textit{average-all} method takes the average of the monitored layer sparsity across all the already executed layers to estimate the dynamic layer sparsity.
This estimated dynamic layer sparsity is then divided by the average sparsity obtained from the sparsity LUT to generate the sparsity coefficient.
Similarly, \textit{last-N} and \textit{last-one} methods follow the same procedure as \textit{average-all}, but estimate the dynamic layer sparsity from the monitored layer sparsity of the last $N$ layers and the last executed layer, respectively.
We evaluated these three approaches in the sparse latency predictor and compared the estimated latency with the measured sparsity obtained from the hardware simulator.
\tabref{tb:rmse} shows the root-mean-square error (RMSE) of latency prediction based on these three approaches, where parameter $N$ of the \textit{last-N} method is tuned through grid search and set to $3$.
We observe that both \textit{average-all} and \textit{last-one} perform similarly, and outperform \textit{last-N}.
As such, we opt to use the \textit{last-one} method as it requires less amount of computation and memory for the averaging operation.

\begin{figure}[t]\centering
    \includegraphics[width=0.49\textwidth]{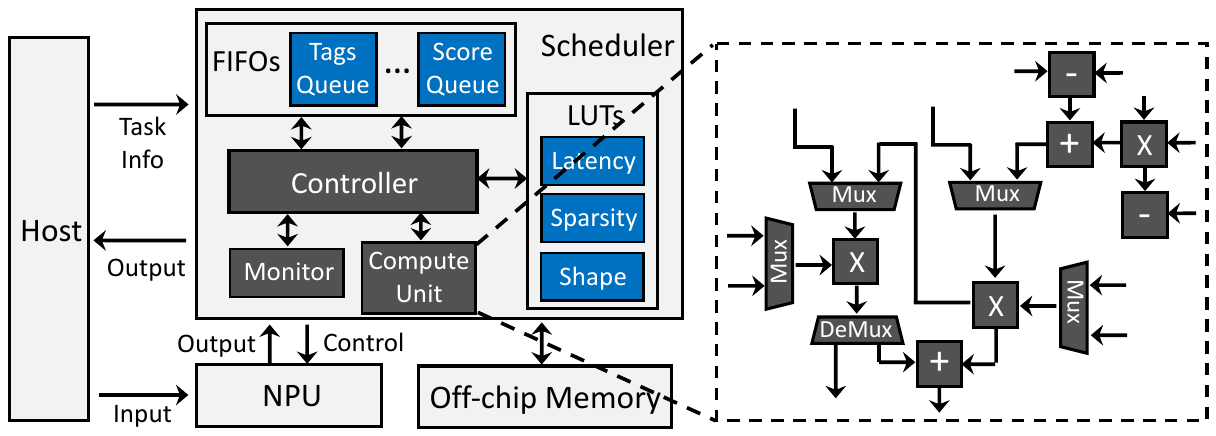}
    \vspace{-2.0mm}
    \caption{Overview of the proposed architecture.}\label{fig:hw_overview}
\end{figure} 

\subsection{Hardware Scheduler Design}\label{subsec:sw_static_scheduler}

\noindent
\subsubsection{Design Overview}\label{subsubsec:hw_scheduler}
\figref{fig:hw_overview} shows an overview of our hardware scheduler's microarchitecture.
The module is situated between the NPU and the host CPU, and is connected to the off-chip memory, if the intermediate results are transferred to it.
We emphasize that the insertion point of our scheduler is dependent on the design of the hardware accelerator.
As we do not impose any restrictions on the accelerator design, we utilize \figref{fig:hw_overview} as an example to demonstrate the use of our hardware scheduler, without loss of generality.

The hardware scheduler consists of a controller, a runtime monitor, a compute unit and multiple FIFOs and LUTs.
The FIFOs are deployed to track the per-request information, such as request tags, scores and latency SLOs.
The depth of FIFOs depends on the maximal number of requests that can be handled by the hardware accelerator.
We set the FIFO depths as configurable parameters in our scheduler.
The latency, sparsity and shape information for each model-pattern pair are cached in three LUTs, which are accessed during the calculation of the sparsity coefficient and score. 
The controller is designed to perform the following tasks: \textit{1)}~receive requests from the static scheduler and forward them into the tag and score FIFOs, \textit{2)}~calculate the sparsity coefficient of the currently running request based on the monitored runtime information (\secref{subsec:sparse_lat_pred}), \textit{3)}~update the score of each request, and \textit{4)}~determine the request with the lowest score to be dispatched next.
The calculation of the sparsity coefficient and the score are both implemented in the shared reconfigurable compute unit, detailed in~\secref{subsebsec:compute_unit}.
Finally, the monitor captures the layer sparsity through a zero-counting circuit.

\begin{figure}[t]\centering
    \includegraphics[width=0.49\textwidth]{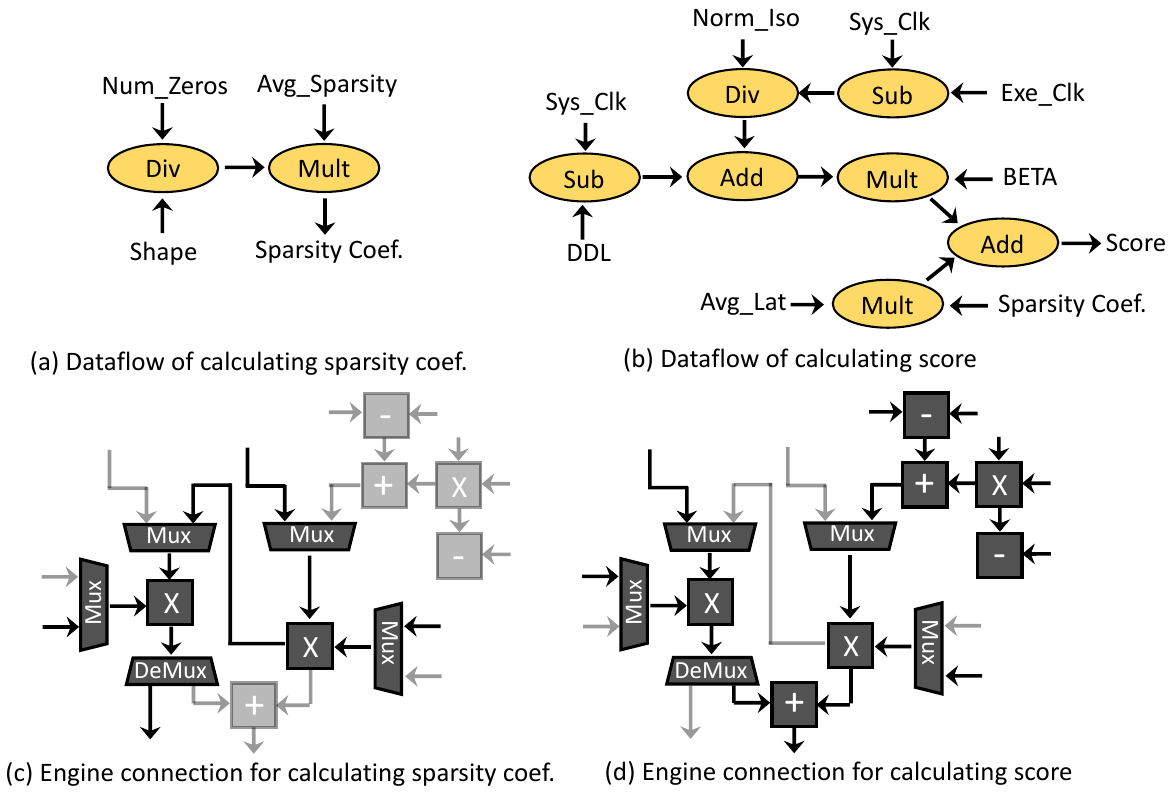}
    \vspace{-2.0mm}
    \caption{Compute engine with adaptable connections. }\label{fig:compute_unit}
\end{figure} 

\subsubsection{Reconfigurable Compute Unit}\label{subsebsec:compute_unit}

To reduce the resource and area overheads of the hardware scheduler,
we propose a reconfigurable compute unit that can be shared for the calculation of both the sparsity coefficient and the request score.
As outlined in Algorithms~\ref{alg:dysta_dynamic_sched} and~\ref{alg:sparse_pred}, the computation comprises, first, calculating the sparsity coefficient $\gamma$ and, then, updating the score of each request based on the estimated sparse latency.
These computations are depicted as computational flows in~\figref{fig:compute_unit}(a) and (b), respectively.
As these computations are performed at different stages of the scheduling process,
we design a reconfigurable compute unit that can be shared by both.

The hardware design of our reconfigurable compute unit is presented on the right-hand side of~\figref{fig:hw_overview}.
We equip the last two multipliers with several multiplexers (Mux) and a de-multiplexer (DeMux) at the input and output ports.
This enables different dataflows to be implemented by reconfiguring the select signals. 
During the computation of the sparsity coefficient,
as the shape is a constant, the division (Div) can be implemented as a multiplication by pre-computing the reciprocal of the shape offline.
Therefore,
the compute unit is configured with the last two multipliers enabled to calculate the sparsity coefficient, as shown in~\figref{fig:compute_unit}(c).
In contrast,
during the computation of scores,
all the arithmetic units are enabled for latency prediction and score aggregation, as shown in~\figref{fig:compute_unit}(d).
The division of normalized isolation time shown in~\figref{fig:compute_unit}(b) is implemented as a multiplication by pre-computing the reciprocal offline.
To further reduce resource consumption,
we adopt half-precision floating-point (FP16) as the data type of the hardware scheduler.

\section{Evaluation}\label{sec:evaluation}
\noindent
\subsection{Experimental Setup}\label{subsec:eval_setup}
\noindent
\textbf{Software Implementation.}
We develop our evaluation and simulation infrastructure using \textit{Python}, as described in~\secref{subsec:evaluate_method}. 
The pre-trained benchmark models (\tabref{tb:sparse_benchmark}) are obtained from \textit{TorchVision} provided by \textit{PyTorch (v3.8.12)} and the \textit{Transformers} library released by \textit{HuggingFace}.
The sparsification process described in~\secref{subsec:model_sparsification} is implemented using \textit{SparseML}~\cite{kurtz2020inducing}.
We adopt the \textit{Sanger}~\cite{sanger201micro} simulator for sparse AttNNs experiments and a custom simulator of \textit{Eyeriss-V2} for sparse CNNs. 
The custom \textit{Eyeriss-V2} simulator is developed based on the validated performance model provided by a third-party code,\footnote{\label{eyeriss_link1}\url{https://github.com/SingularityKChen/dl_accelerator}} to ensure its correctness.
The multi-AttNN workloads target the real scenarios on mobile phones with a mix of machine translation and question \& answering as shown in~\tabref{tb:sparse_benchmark}.
The Multi-CNNs constitute a mix of visual perception and hand tracking, which are common workloads in AR/VR~\cite{xrbench2023mlsys}, robotics~\cite{jang2019mnnfast} and data centers~\cite{venieris2022multi}.

\noindent
\textbf{Hardware Implementation.}
We implement our hardware scheduler using \textit{SystemVerilog}.
The scheduler is clocked at $200$ MHz.
To make a consistent evaluation as \textit{Eyeriss-V2} design\textsuperscript{\ref{eyeriss_link1}} on FPGA, Xilinx Vivado 2019.1 is used for synthesis and implementation.
To support large CNN models such as \textit{ResNet-50} and \textit{VGG-16},
we increase the global buffer of input activations in \textit{Eyeriss-V2} from $1.5$KB to $2.5$KB.
 The other hardware configurations of both \textit{Sanger} and \textit{Eyeriss-V2} remain consistent with their original papers.

\begin{table}[t]
\centering
\caption{Comparison of scheduling approaches.}
\vspace{2.0 mm}
\label{tb:e2e_compare}
\setlength\tabcolsep{1pt} 
\scalebox{0.77}{
\begin{tabular}{L{2.25cm}|C{1.7cm}|C{1.8cm}|C{1.7cm}|C{1.8cm}}
\toprule \midrule
 & \multicolumn{2}{c|}{\bf Multi-AttNNs} & \multicolumn{2}{c}{\bf Multi-CNNs} \\ \cmidrule{2-5}
 & {\bf ANTT $\downarrow$}& {\bf Violation Rate [\%] $\downarrow$} & {\bf ANTT $\downarrow$}& {\bf Violation Rate [\%] $\downarrow$}\\ \midrule
 {\bf FCFS} & 18.9 & 55.1 & 11.4 & 23.1 \\ \midrule
 {\bf SJF} & 5.0 & 15.2 & 2.6 & 3.4 \\ \midrule
 {\bf SDRM$^3$~\cite{kim2022sdrm3}} &18.9 & 63.3 & 9.3 & 33.7\\ \midrule
 {\bf PREMA~\cite{choi2020prema}} & 5.4 & 15.3 & 3.0 & 3.2 \\ \midrule
 {\bf Planaria~\cite{ghodrati2020planaria}} &16.0 & 6.8 & 4.2 & 2.1\\ \midrule
 {\bf Dysta (Ours)} & {\bf 4.7} & {\bf 5.1} & {\bf 2.5} & {\bf 2.0} \\ \midrule
\bottomrule
\end{tabular}}
\end{table}

\begin{figure}[t]
    \vspace{-0cm}
    \centering 
    \subfigure{\includegraphics[width=0.49\textwidth,trim={0cm 0cm 0cm 0cm},clip]{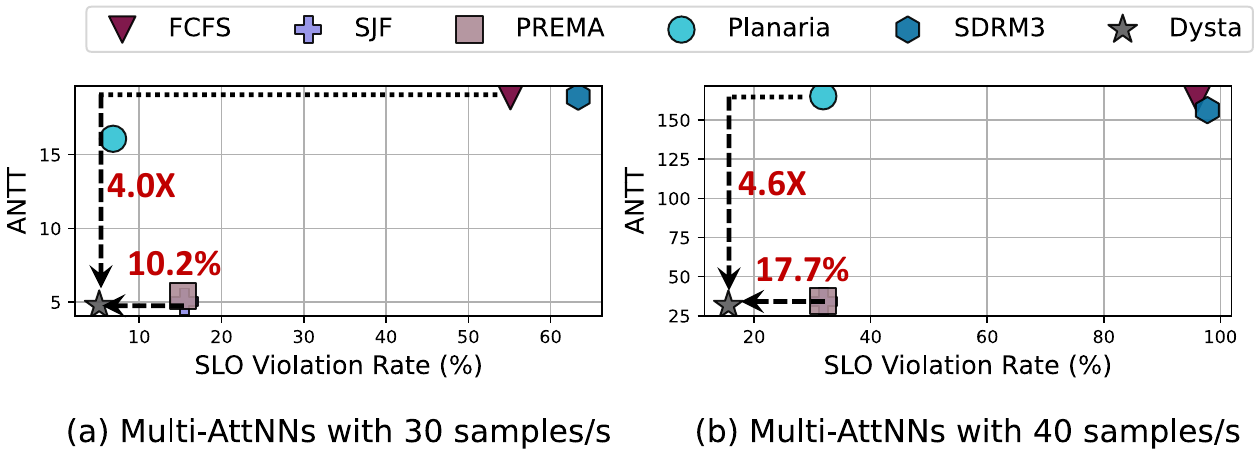}} 
    \subfigure{\includegraphics[width=0.49\textwidth,trim={0cm 0cm 0cm 0cm},clip]{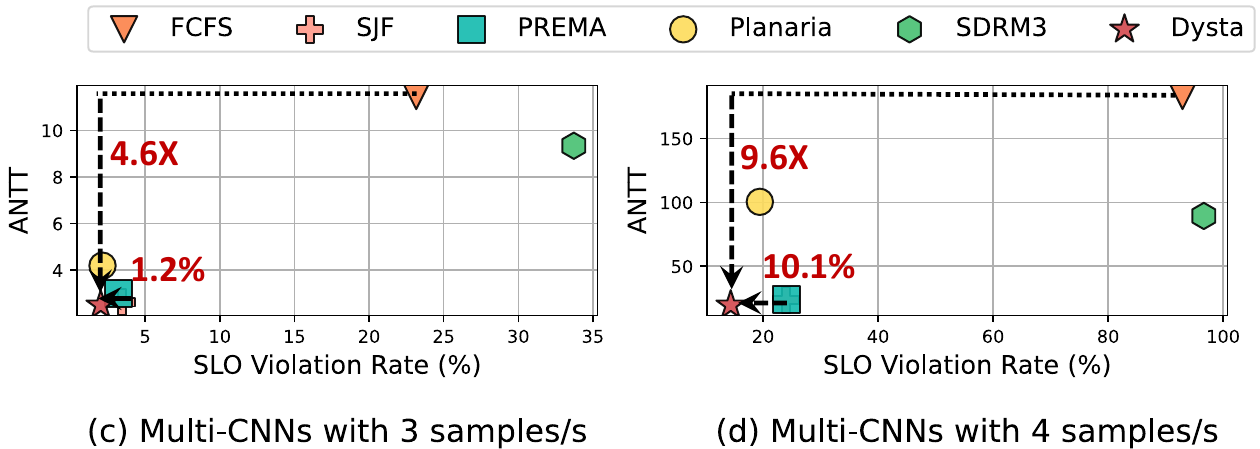}} 
    \caption{SLO violation rate and ANTT trade-off.}
    \label{fig:trade-off}
\end{figure}

\noindent
\textbf{Baselines.}
We compare against status-quo scheduling approaches, including \textit{i)}~First-Come~First-Served (FCFS) and \textit{ii)}~Shortest-Job First (SJF), and the state-of-the-art multi-DNN schedulers: \textit{iii)}~PREMA~\cite{choi2020prema}, \textit{iv)}~Planaria~\cite{ghodrati2020planaria}, and \textit{v)}~SDRM\textsuperscript{3}~\cite{kim2022sdrm3}. 
To improve the performance of PREMA at the beginning of the scheduling,
we modify the selection criterion (line~9 of the PREMA scheduling algorithm presented in~\cite{choi2020prema}) to $Token_i \geq Threshold$, as opposed to the original condition of $Token_i > Threshold$.
For Planaria, we set the resource requirement estimate (line~40 in Algorithm~1 in~\cite{ghodrati2020planaria}) to 1 for all tasks, since our target accelerators are time-shared, without spatial co-location of multiple tasks.
For SDRM$^3$, we express \texttt{MapScore} (Eq.~(5) in~\cite{kim2022sdrm3}) as the weighted sum of \textit{Urgency} and \textit{Fairness}. We further set the weight \textit{Pref} to 1, since we target only one accelerator at a time, and tune parameter $\alpha$ 
following SDRM$^3$'s optimization methodology.

\noindent
\textbf{Metrics.}
To assess the performance of different scheduling approaches,
we utilize three metrics: average normalized turnaround time (ANTT), SLO violation rate and system throughput (STP).
Specifically, for a multi-DNN workload consisting of $N$ requests,
we define ANTT as $\frac{1}{N} \sum_{n=1}^{N} \frac{T_{i}^{\text{Multi}}}{T_{i}^{\text{Isol.}}}$ where $T_{i}^{\text{Multi}}$ is the measured execution time under multi-tenancy and $T_{i}^{\text{Isol.}}$ is the uninterrupted isolated execution time of the target task.
The SLO violation rate is calculated as $\frac{N^{\text{viol}}}{N}$ where $N^{\text{viol}}$ represents the number of violated tasks under certain latency SLOs and request arrival rates.
Following the experimental setup of~\cite{choi2020prema},
we set the SLO as $T_{i}^{\text{Isol.}} \times M^{\text{lat}}_{\text{slo}}$ where $M^{\text{lat}}_{\text{slo}}$ refers to the SLO multiplier used to control the stringency of the latency constraints.
We set the total number of requests in each workload as $1000$ to ensure the reliability of our results.
For each metric,
we evaluate using five random seeds
and report the average.

\begin{figure}[t]
    \vspace{-0cm}
    \centering
    \subfigure[Optimization breakdown in multi-AttNNs.]{\includegraphics[width=0.46\textwidth,trim={0cm 0cm 0cm 0cm},clip]{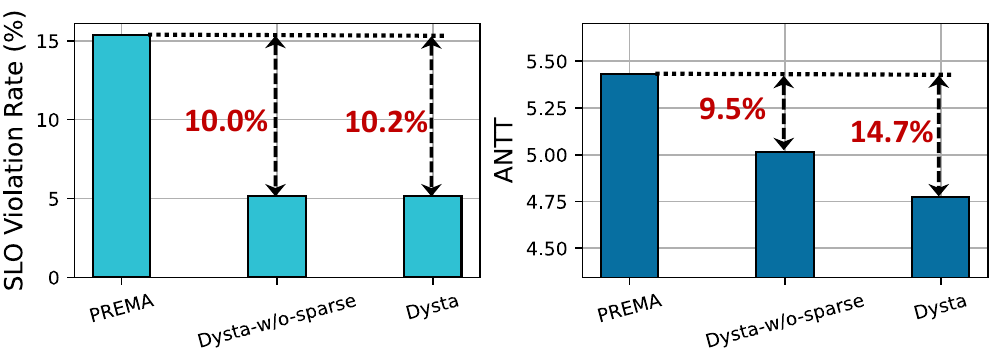}} 

    \subfigure[Optimization breakdown in multi-CNNs.]{\includegraphics[width=0.46\textwidth,trim={0cm 0cm 0cm 0cm},clip]{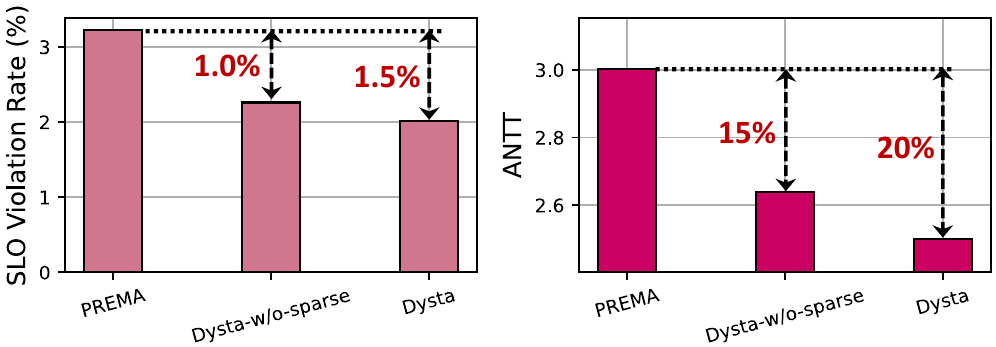}}
    \vspace{-3mm}
    \caption{Optimization breakdown.}
    \label{fig:effect_sparsity}
\end{figure}

\subsection{End-to-End Performance Comparison}\label{subsec:e2e_eval}

We evaluate the performance of different scheduling approaches in two multi-tenant scenarios: multi-AttNNs and multi-CNNs.
To generate the multi-AttNN and multi-CNN workloads, we randomly sample from AttNNs and CNNs listed in~\tabref{tb:sparse_benchmark}, respectively.
Following the \textit{MLPerf} standard~\cite{mlperf_inf2020isca},
we generate the request arrival times for each workload using a Poisson distribution.
Given the computational capacity of \textit{Sanger} and \textit{Eyeriss-V2},
we set the arrival rate of the multi-AttNN workload as $30$ samples/s and the multi-CNN workload as $3$ samples/s.
The latency SLO multiplier is configured as $10\times$ for both multi-AttNN and  multi-CNN workloads.
\tabref{tb:e2e_compare} presents the violation rate and ANTT of different scheduling approaches.
We note that our proposed approach achieves similar performance gains under workloads with different arrival rates and latency SLO multipliers, as demonstrated in~\secref{sub_sec:robust}.

\begin{figure}[t]
    \centering
    \subfigure[Multi-AttNNs with arrival rate of 30 samples/s.]{\includegraphics[width=0.5\textwidth,trim={0cm 0cm 0cm 0cm},clip]{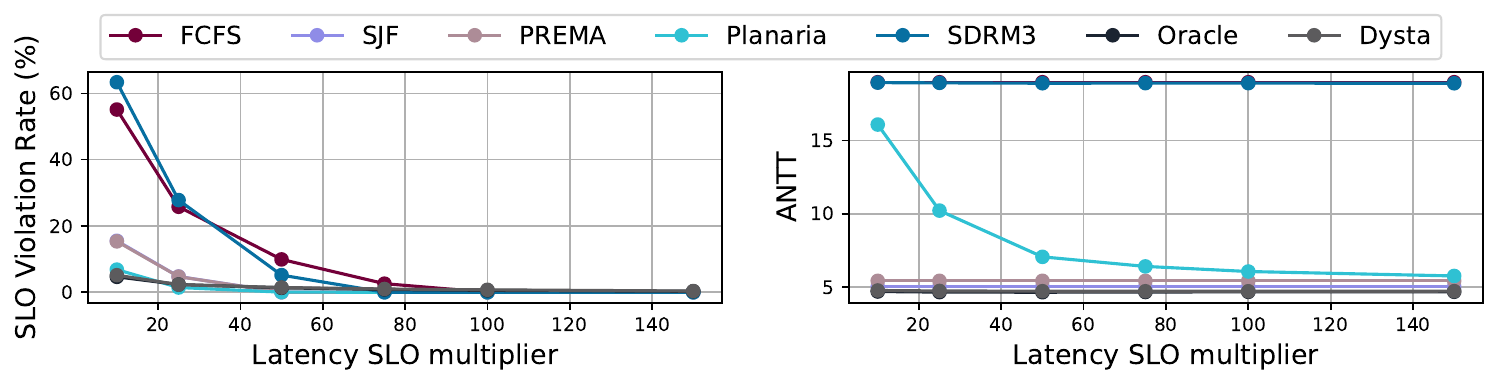}} 
    \subfigure[Multi-AttNNs with arrival rate of 40 samples/s.]{\includegraphics[width=0.5\textwidth,trim={0cm 0cm 0cm 0cm},clip]{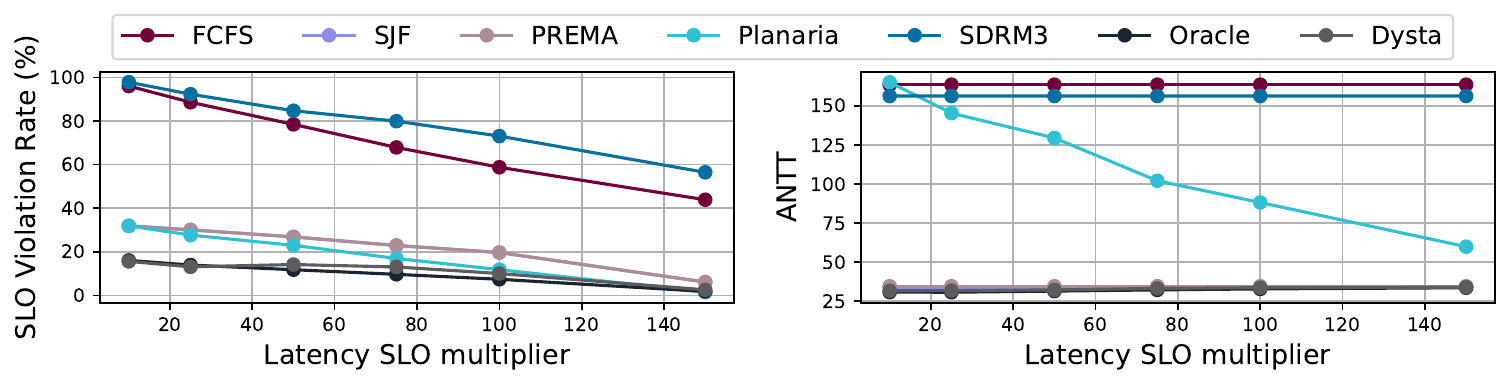}} 
    \subfigure[Multi-CNNs with arrival rate of 3 samples/s.]{\includegraphics[width=0.5\textwidth,trim={0cm 0cm 0cm 0cm},clip]{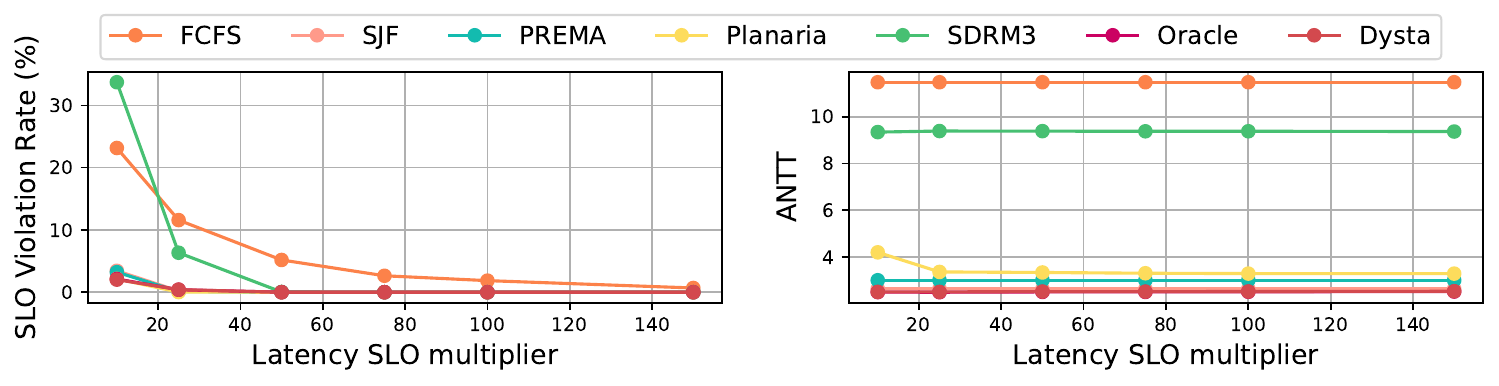}}
    \subfigure[Multi-CNNs with arrival rate of 4 samples/s.]{\includegraphics[width=0.5\textwidth,trim={0cm 0cm 0cm 0cm},clip]{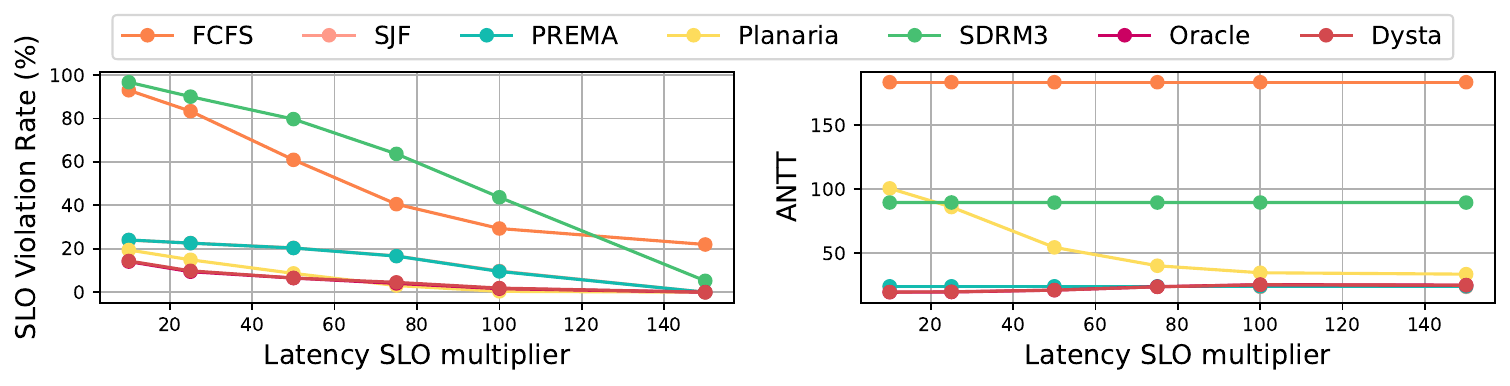}}
    \vspace{-3mm}
    \caption{Evaluation across different latency SLOs.}
    \label{fig:slos_eval}
    \vspace{-1mm}
\end{figure}

\begin{figure*}[!htb]
    \vspace{-0cm}
    \centering
    \subfigure[SLO violation rate, throughput  and ANTT of multi-AttNNs.]{\includegraphics[width=0.96\textwidth,trim={0cm 0cm 0cm 0cm},clip]{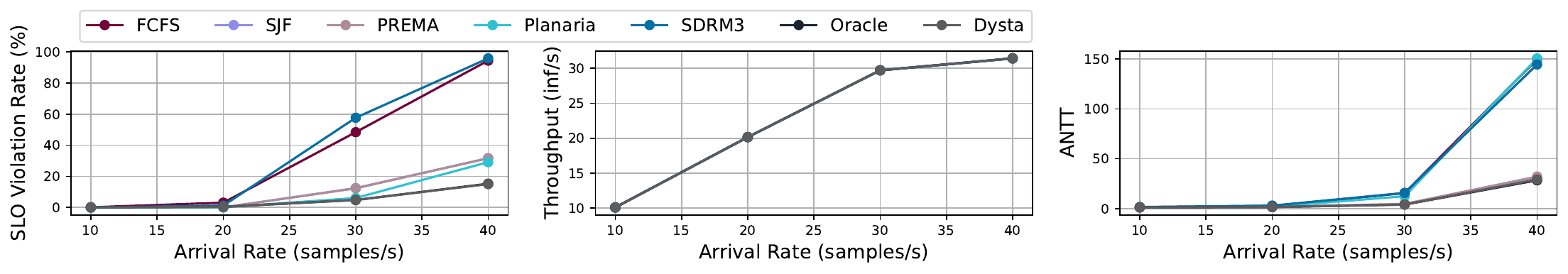}} 
    \subfigure[SLO violation rate, throughput and ANTT of multi-CNNs.]{\includegraphics[width=0.96\textwidth,trim={0cm 0cm 0cm 0cm},clip]{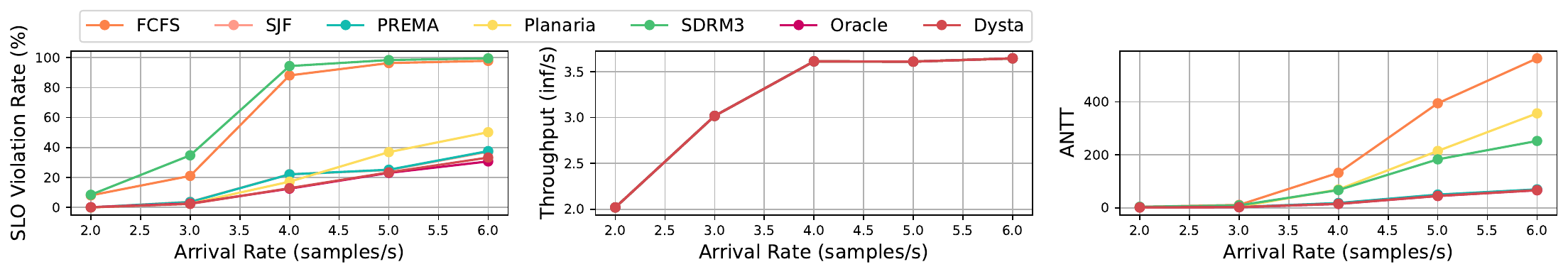}}
    \vspace{-3mm}
    \caption{Evaluation across different arrival rates on multi-AttNN and multi-CNN workloads.}\label{fig:arrival_rate_eval}
    \vspace{-2mm}
\end{figure*}

\begin{figure}[t]\centering
    \includegraphics[width=0.49\textwidth]{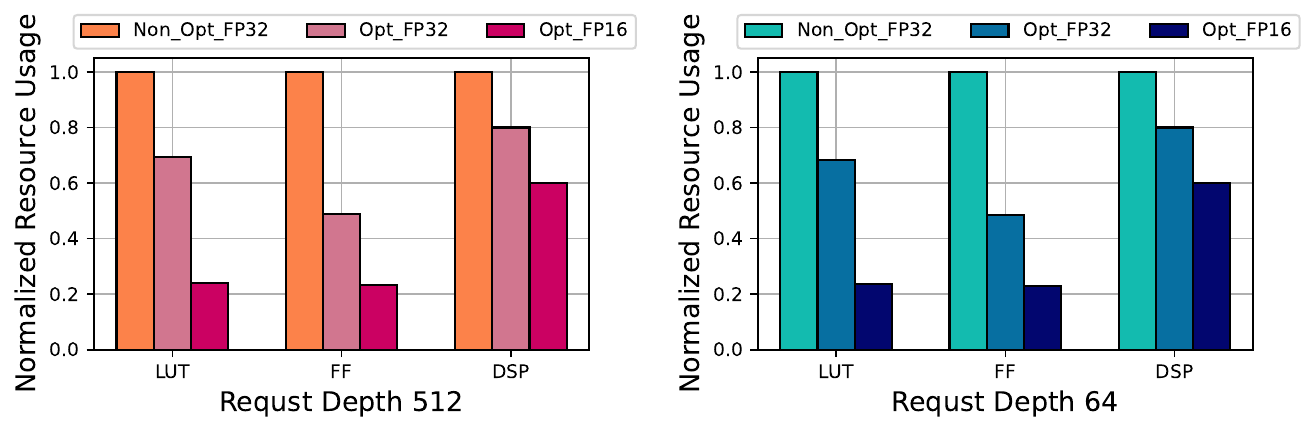}
    \vspace{-6.0mm}
    \caption{Resource usage with different optimizations.}\label{fig:hw_effect}
\end{figure} 

As seen from~\tabref{tb:e2e_compare},
our \textit{\sys} scheduler outperforms all the previous approaches.
In contrast to previous approaches, such as \textit{PREMA} and \textit{Planaria}, which only perform well on either violation rate or ANTT,
our proposed scheduler excels in both metrics compared to the traditional heuristic \textit{SJF} method.
Comparing with the SOTA \textit{Planaria},
our approach achieves a reduction of $1.7$\% in the violation rate while decreasing ANTT by $3.4\times$ in multi-AttNN workloads.
\textit{\sys} also reduces the ANTT by nearly $2\times$ in multi-CNNs compared with \textit{SJF} while achieving similar violation rates.

To visualize the improved trade-off achieved by \textit{\sys},
we present a 2D plot with ANTT on the y-axis and violation rate on the x-axis, as shown in~\figref{fig:trade-off}. 
The multi-AttNN workloads are evaluated at arrival rates of $30$ and $40$ samples/s, while the multi-CNN workloads are evaluated at arrival rates of $3$ and $4$ samples/s.
The SLO configuration remains the same as in~\tabref{tb:e2e_compare}.
As observed from~\figref{fig:trade-off},
\textit{\sys} is located at the lower left corner with the lowest violation rate and ANTT compared with other approaches, demonstrating its effectiveness in achieving an improved trade-off and better matching the multi-objective nature of multi-DNN systems.

\subsection{Optimization Breakdown}\label{subsec:lat_pred}
To investigate the gain introduced by each proposed technique,
we provide optimization breakdowns for both multi-AttNN and multi-CNN workloads, as shown in~\figref{fig:effect_sparsity}.
Two baselines are evaluated to compare with \textit{\sys}: \textit{1)}~\textit{PREMA}, a representative SOTA multi-DNN scheduling approach, and \textit{2)}~\textit{Dysta-w/o-sparse}, an ablated variant of \textit{Dysta} with dynamic hardware scheduler and sparsity-aware support disabled.
By comparing \textit{Dysta-w/o-sparse} against \textit{PREMA},
a clear reduction in the violation rate and ANTT can be observed on both multi-AttNN and multi-CNN workloads,
demonstrating the effectiveness of our static score-based scheduling approach.
Comparing \textit{Dysta} to \textit{Dysta-w/o-sparse},
we can see a clear ANTT drop on both multi-AttNN and multi-CNN workloads by adopting the dynamic hardware component.
In general, we observe that the dynamic hardware scheduler has a lesser impact on the violation rate than ANTT.
This can be attributed to the significant correlation between task violations and latency SLO objectives.
With loose SLO objectives, the latency estimated by the static scheduler is accurate enough to prevent task violation.
Consequently, no further significant improvement is achieved by the dynamic hardware component.

\subsection{Robustness Evaluation}\label{sub_sec:robust}
To validate the robustness of our scheduling approach,
we conduct experiments using workloads generated with different hyperparameters, \textit{i.e.}~latency SLOs and arrival rates.

\subsubsection{Robustness across Different Latency SLOs}\label{subsub_sec:slo_eval}
To evaluate the performance with different SLO requirements,  we vary the latency SLO multiplier between $10\times$ and $150\times$ for both multi-AttNN and multi-CNN workloads.
We evaluate two arrival rates for each workload,
which are configured as $30$ and $40$ samples/s for multi-AttNNs and $3$ and $4$ samples/s for multi-CNNs.
\figref{fig:slos_eval} shows the results of the violation rate and ANTT.
As the latency SLO multiplier increases,
both ANTT and SLO violations exhibit a declining trend due to the relaxed latency constraint.
In multi-AttNN workloads,
\textit{\sys} achieves the lowest violation rate and ANTT across different latency SLOs.
This trend can be observed in both arrival rates ($30$ and $40$ samples/s), demonstrating the advantage and robustness of our approach against different levels of latency SLO requirements.
\textit{\sys} also closely matches the performance of the Oracle scheduler in ANTT and violation rate.
The same conclusion can be drawn for multi-CNNs, where \textit{\sys} consistently outperforms other approaches in both violation rate and ANTT across different latency SLO multipliers.
This also demonstrates the robustness of our approach under different families of models.

\subsubsection{Robustness across Different Arrival Rates}\label{subsub_sec:ar_eval}

To evaluate the performance across traffic levels, 
we vary the request arrival rate between $10 \thicksim 40$ samples/s for multi-AttNNs, and $2 \thicksim 6$ samples/s for multi-CNNs, based on the computational capacity of the target hardware. 
To eliminate the impact of latency SLOs,
we set $M^{\text{lat}}_{\text{slo}}$ to $10\times$ for both workloads.
\figref{fig:arrival_rate_eval} shows the violation rate, system throughput and ANTT of different scheduling methods across different arrival rates.
All three metrics exhibit an upward trend as the arrival rate increases.
The change in throughput remains the same across different scheduling methods as it is dependent on the computational capacity of the hardware.
Under different arrival rates,
\textit{\sys} keeps outperforming other schedulers in both violation rate and ANTT, with close performance to Oracle.
The gain becomes more prominent while increasing the arrival rate,
demonstrating the robustness of our method under heavier traffic.

\subsection{Hardware Overhead}\label{subsec:hw_eval}

To evaluate the hardware overhead introduced by the proposed hardware scheduler,
we compare its resource utilization against \textit{Eyeriss-V2}.
We adopt an open-source third-party hardware design\footnote{\label{eyerissv2_link}\url{https://github.com/karthisugumar/CSE240D-Hierarchical_Mesh_NoC-Eyeriss_v2}} to get the resource consumption of \textit{Eyeriss-V2}.
Both the hardware scheduler and accelerator are clocked at $200$~MHz, targeting the Xilinx Zynq ZU7EV FPGA board.

To demonstrate the resource reduction provided by the reconfigurable compute unit and FP16 representation,
we evaluate the resource usage of three hardware designs with different optimizations applied: \textit{1)}~$Non\_Opt\_FP32$, which refers to the native implementation with separate compute units and 32-bit floating-point (FP32), \textit{2)}~$Opt\_FP32$ that adopts reconfigurable compute unit, and \textit{3)}~$Opt\_FP16$ with both reconfigurable compute unit and FP16 applied.
To validate the effectiveness of our optimizations across different FIFO depths (\secref{subsubsec:hw_scheduler}),
we instantiate two designs with FIFO depth of $512$ and $64$, respectively.
As can be seen in~\figref{fig:hw_effect},
the reconfigurable compute unit brings a significant reduction in LUT, register (Flip-Flops (FF)) and DSP resources.
These reductions come from the savings of the compute unit, logic and memory required for calculating the sparsity coefficient.
By comparing $Opt\_FP16$ with $Opt\_FP32$,
we can also see significant reductions in all three types of resources.
For both optimizations, a similar reduction trend can be observed for both FIFO depths, demonstrating the effectiveness of our optimizations across different FIFO depths.

\tabref{tb:hw_overhead} presents the hardware overhead of our approach compared to \textit{Eyeriss-V2}.
Given the computational capacity of \textit{Eyeriss-V2},
we set the FIFO depth to $64$.
Our hardware scheduler consumes a minimal amount of $0.55$\% of LUTs, $1.5$\% of DSPs and $0.35$\% of on-chip RAM, which is negligible compared to the overall resource usage.
The significant gains achieved in ANTT and violation rate with such a low hardware cost demonstrates the effectiveness of our approach.

\begin{table}[t]
    \centering
    \caption{Resource overhead of \textit{\sys} \mbox{scheduler}.}
    \vspace{0.2cm}
    \label{tb:hw_overhead}
    \resizebox{0.37\textwidth}{!}{
        \setlength{\tabcolsep}{3pt}
        \begin{tabular}{@{} l l l l @{}}
            \toprule
            \textbf{Module} & \textbf{LUTs} & \textbf{DSPs} & \textbf{On-Chip RAM} \\ 
            
            \midrule 
            \textbf{Eyeriss-V2} & 99168 & 194 & 140~KB \\
            \textbf{Scheduler} & 553  & 3 & 0.5~KB \\
            \textbf{\sys-Eyeriss-V2} & 99721  & 196 & 140.5~KB \\
            
            \midrule
            \textbf{Total Overhead} & 0.55\%  & 1.5\% & 0.35\% \\
            \bottomrule
        \end{tabular}
    } 
\end{table}

\section{Related Work}\label{sec:background}

\noindent
\textbf{Multi-DNN Hardware Accelerators.}
Hardware architectures for multi-DNN workloads can be taxonomized based on their flexibility into: \textit{1)}~fixed-purpose~\cite{fcnnx2018fpl,kwon2021heterogeneous}, where the multi-DNN workloads are know \textit{a piori}; and \textit{2)}~workload-agnostic accelerators~\cite{choi2020prema,aimt2020isca,ghodrati2020planaria,dataflow_mirroring2021dac}. \textit{Fixed-purpose designs} opt to instantiate multiple heterogeneous compute engines~\cite{fcnnx2018fpl,kwon2021heterogeneous,multidnn_codesign2020dac} that operate in parallel and apply design-time optimizations for the given multi-DNN workload. Prominent optimizations include highly customized compute engines and static scheduling~\cite{fcnnx2018fpl,kwon2021heterogeneous}, both tailored to the target workloads, or co-design of the set of models and the hardware~\cite{multidnn_codesign2020dac}. \textit{Workload-agnostic designs} have either augmented existing accelerators with the necessary hardware components to support preemptive time-multiplexing~\cite{choi2020prema,aimt2020isca} or proposed highly flexible architectures that enable the spatial co-location of multiple DNNs through dynamic resource allocation~\cite{ghodrati2020planaria,dataflow_mirroring2021dac}.
Finally, another line of work has investigated different trade-offs of multi-DNN hardware architectures through design space exploration~\cite{multidnn2020les,magma2022hpca,serv_multidnn_fpga2022tc} and interference-aware performance modeling~\cite{infer2020fpt}.
In this work, we adopt preemptive time-multiplexing architectures, which constitute the most widespread deployed design. This allows \textit{\sys}'s techniques to improve performance with minimal hardware modifications, leading to wider generality and impact.

\noindent
\textbf{Multi-DNN Schedulers.}
Schedulers for multi-DNN workloads have so far adopted two main approaches: \textit{1)}~temporal and \textit{2)}~spatio-temporal scheduling. In the first approach, a single model occupies the compute engine of the target accelerator at each time instant. The optimizations have focused on preemptive scheduling schemes~\cite{choi2020prema}, inter-DNN pipelining~\cite{aimt2020isca,layerweaver2021hpca} that co-schedules compute- and memory-bound DNNs for higher utilization of both the computational and memory resources, and memory-aware scheduling to reduce memory swapping between models~\cite{masa2021percom}. In the second approach, multiple models are processed in parallel by different parts of the accelerator. In this case, the scheduler decides both which DNNs to dispatch and the resource allocation among them, statically~\cite{fcnnx2018fpl, kwon2021heterogeneous, magma2022hpca} or dynamically~\cite{ghodrati2020planaria, dataflow_mirroring2021dac, kim2022sdrm3}.

Despite the progress, as shown in~\tabref{tb:comp_prev_work}, existing methods are constrained by two main limitations. First, by neglecting the presence of model sparsity, they leave untapped optimization opportunities. Second, each scheduler tends to work well on a single performance metric, \textit{e.g.}~either SLO violation rate or ANTT, while excessively penalizing the other. Instead, through its sparsity-aware, fine-grained scheduling algorithm and the low-cost hardware enhancement, \textit{\sys} pushes the boundaries of the attainable performance, while consistently yielding a better trade-off between SLO violations and ANTT.

\noindent
\textbf{Multi-DNN Benchmarks.}
So far, progress has been made primarily towards benchmark suites for single-DNN execution, with prominent efforts such as AI Benchmark for mobile devices~\cite{ai_benchmark2019iccvw} and variants of \textit{MLPerf}~\cite{mlperf_inf2020isca, mlperf_mobile2022mlsys}. Closer to our setting lies XRBench~\cite{xrbench2023mlsys}, a domain-specific benchmark suite of multi-DNN workloads for AR/VR applications. Nonetheless, there is still a lack of standardization in the evaluation of more generic multi-DNN systems, and their performance under sparse DNNs. We aim to pave the way towards bridging this gap through the open-source benchmark and evaluation infrastructure that we introduced in this work.

\section{Conclusion}\label{sec:conlcusion}

The increasing demand for running multiple DNNs in parallel and the prevalence of sparsity across different DNN models have led to the emergence of sparse multi-DNN workloads.
By identifying the optimization opportunities in sparse multi-DNN workloads,
we propose a novel bi-level dynamic and static scheduler 
that utilizes sparsity dynamicity and pattern information for better scheduling.
Coupled with an efficient hardware scheduler and sparse latency predictor,
our proposed approach achieves up to $10$\% fewer violations and nearly $4\times$ lower average normalized turnaround time compared to the state-of-the-art methods, while incurring negligible hardware cost. 
To facilitate future development in this area,
we will open-source all the benchmarks and code upon the paper acceptance.
We believe our contributions will attract further research attention to study sparse multi-DNN workloads.

%
%
%
%
%

\appendix
\section{Artifact Appendix}

\subsection{Abstract}
This Appendix summarizes the necessary information and instructions to reproduce our experimental results.
Our artifacts mainly contain \textit{1)}~public sparse multi-DNN benchmark, \textit{2)}~simulation-based evaluation infrastructure for multi-DNN scheduling and \textit{3)}~hardware prototype of the proposed dynamic scheduler. 
The profiling results and algorithmic performance can be reproduced by running our Python and PyTorch programs.
The hardware resource consumption can be obtained by running Synthesis and Implementation on Vivado using our RTL code and constraint files.
To facilitate reviewers and readers with quickly reproducing our results, we provide Bash scripts to re-generate all the figures and tables of our paper.

\subsection{Artifact check-list (meta-information)}

\begin{itemize}
  \item {\bf Algorithm: } \textit{\sys} scheduler, a sparsity-aware dynamic and static scheduling algorithm for sparse multi-DNN workloads.
  \item {\bf Program: } Python, PyTorch, SparseML, Verilog HDL.
  \item {\bf Model: } For multi-CNN workloads, we adopt four popular CNN models, namely \textit{SSD}~\cite{liu2016ssd}, \textit{ResNet-50}~\cite{he2016deep}, \textit{VGG-16}~\cite{simonyan2015very}, and \textit{MobileNet}~\cite{mobilenet2017arxiv}. For multi-\mbox{AttNNs} workloads, our benchmark includes three commonly used language models: \textit{BERT}~\cite{devlin2018bert}, \textit{BART}~\cite{lewis2019bart}, and \textit{GPT-2}~\cite{radford2019language}.
  \item {\bf Data set: } ImageNet~\cite{deng2009imagenet}, ExDark~\cite{loh2019getting}, DarkFace~\cite{Chen2018Retinex}, and COCO~\cite{lin2014microsoft} are used for vision tasks. GLUE~\cite{wang2018glue} and SQUAD~\cite{rajpurkar2016squad} are used for language tasks.
  \item {\bf Run-time environment: } Ubuntu~20.04, CUDA SDK~11.3 or higher.
  \item {\bf Hardware: } Nvidia RTX 2080 GPU, Intel Xeon Gold 6154 CPU.
  \item {\bf Metrics: } Latency service-level objective (SLO) violation rate, average normalized turnaround time (ANTT) and hardware resource consumption.
  \item {\bf Experiments: } Bash scripts and detailed instructions are provided to run experiments.
  \item {\bf How much disk space required (approximately)?: } 20~GB.
  \item {\bf How much time is needed to prepare workflow (approximately)?: } $1\sim2$ hours.
  \item {\bf How much time is needed to complete experiments (approximately)?: } Scheduling results need 3 hours. Resource consumption requires 5 hours. 
  \item {\bf Publicly available?: } Yes.
  \item {\bf Code licenses (if publicly available)?: } Yes.
  \item {\bf Archived (provide DOI)?: } 10.5281/zenodo.8252767
\end{itemize}

\subsection{Description}
\subsubsection{How to access}\label{subsec:docker}
You can access our codebase from the link: \url{https://github.com/SamsungLabs/Sparse-Multi-DNN-Scheduling}.
Instructions on how to download the Docker image are provided in {\path{README.md}}.

\subsubsection{Hardware dependencies} A GPU server is required to run the training of sparse models.
A CPU server is needed to run simulation, synthesis and place-\&-route. 

\subsubsection{Software dependencies}  Vivado Design Suite $2019.2$, PyTorch $1.11.0$, CUDA SDK~$11.3$ or higher, Python~$3.8$ or higher. Other dependencies are described in {\path{README.md}}.

\subsection{Installation}

We provide a detailed installation guide in the {\path{README.md}} of the root directory.
\subsection{Experiment workflow}\label{subsec:exp_flow}
To evaluate our artifacts, perform the following steps:
\begin{itemize}
    \item Follow the instructions of experimental setup in the root directory to install software dependencies. Or Download the Docker image provided in~\secref{subsec:docker}. Instructions on loading and creating the Docker container are provided in {\path{README.md}}.
    \item Download the necessary vision and language datasets. (Optional if using Docker)
    \item Generate CSV files using the hardware simulator in {\path{hw_simulator}}. (Optional if using Docker)
    \item Follow \path{INST_RPRODUCE.md} to run experiments using the scripts we provide.
\end{itemize}

\subsection{Evaluation and expected results}
We provide Bash scripts to generate all the figures and tables related to scheduling performance and resource consumption.
Detailed instructions on how to run scripts are described in \path{INST_RPRODUCE.md}.
As running all the experiments requires a few hundred GPU/CPU hours, to facilitate the artifact evaluation,
we refer to the following key results that can be obtained within a reasonable time:
\begin{itemize}
    \item Profiling results of runtime variance while simulating different sparse models on CNN and AttNN accelerators. Following \path{INST_RPRODUCE.md} can reproduce \figref{fig:sparsity_bert_sanger}, \figref{fig:sparsity_layer_relu}, \figref{fig:sparsity_pattern}, and \tabref{tb:network_sparse}. 
    \item ANTT and latency SLO violation rate trade-off as shown in~\tabref{tb:e2e_compare}, \figref{fig:trade-off} and~\figref{fig:effect_sparsity}. Following \path{INST_RPRODUCE.md} to run  Bash scripts under folder \path{dysta_scheduler/script}.
    \item Stress test of scheduling approaches under different latency SLOs and arrival rates as shown in~\figref{fig:slos_eval} and~\figref{fig:arrival_rate_eval}. Following \path{INST_RPRODUCE.md} to run Bash scripts under folder \path{dysta_scheduler/script}.
    \item Hardware resource consumption under different hardware optimizations. Following \path{INST_RPRODUCE.md} to download Vivado reports and run the script under \path{/hw_design/draw_fig_hw_opt.sh}
\end{itemize}

\subsection{Methodology}

Submission, reviewing and badging methodology:

\begin{itemize}
  \item \url{https://www.acm.org/publications/policies/artifact-review-badging}
  \item \url{http://cTuning.org/ae/submission-20201122.html}
  \item \url{http://cTuning.org/ae/reviewing-20201122.html}
\end{itemize}



\bibliographystyle{ACM-Reference-Format}
\bibliography{micro23_spar_dysta}


\end{document}